\documentclass [11pt, proquest] {uwthesis}[2020/02/24]

\usepackage{alltt}  

\usepackage{enumitem}
\usepackage{hyperref}
\hypersetup{              % configure hyperref
    pdfdisplaydoctitle=true, % For Accessibility
    bookmarksnumbered=true,
    colorlinks=true,        % false: boxed links; true: colored links
    citecolor=black,        % color of links to bibliography
    filecolor=black,        % color of file links
    linkcolor=black,        % color of internal links
    urlcolor=linkColor,     % color of external links
    bookmarksopen=true,
    pdfpagemode=UseNone,   % no bookmarks shown when file opened
    breaklinks=true,
    pdfstartview={FitH},
}
\usepackage{cleveref}
\usepackage{graphicx}
\usepackage{enumitem}
\usepackage{listings}
\usepackage{enumitem}
\setenumerate[0]{label=(\roman*)}
\setenumerate[1]{label=(\alph*)}
\usepackage{amsfonts}
\usepackage{longtable}
\usepackage{subcaption}

\begin{document}
\definecolor{linkColor}{RGB}{6,125,233}

 \newenvironment{enumeratequestions}{
    \begin{enumerate}
        \setlength{\itemsep}{12pt}
}{
    \end{enumerate}
}

\prelimpages
 
\Title{Does This Even Matter in the Real World?\\Real World Problems in Foundational Theory Courses}
\Author{Anna Kuznetsova}
\Year{2026}
\Program{Paul G. Allen School of Computer Science}

\Chair{Associate Teaching Professor, Robbie Weber}{Associate Teaching Professor}{Department of Chair}

\copyrightpage

\titlepage  

%
% ----- abstract
%

\setcounter{page}{-1}
\abstract{
Discrete mathematics and probability theory contain foundational material for computer scientists. Despite their importance, instructors often worry that students will find these courses to be too abstract and seemingly disconnected from their future careers. For this research project, we introduced homework questions throughout our introductory theory courses based on real world applications of the course content. Areas of application included a court case, code correctness, and machine learning ethics. We surveyed students at the beginning and end of the term on their attitudes toward the relevance of the course material. Our results, surprisingly, indicate that a small minority of students (less than  $7\%$) expected the material to be irrelevant to them at the start of the term, and a similarly small number believed that at the end of the term. Our surveys and qualitative feedback also indicate students enjoyed having the problems and wanted them to continue being offered in future iterations of the courses.
}
 
%
% ----- contents & etc.
%
\tableofcontents
\listoffigures
\listoftables
 
%
% ----- acknowledgments
%
\acknowledgments{% \vskip2pc
  % {\narrower\noindent
  The author wishes to express sincere appreciation to
  University of Washington, where she has had the opportunity
  to work under the mentorship of Associate Teaching Professor, Robbie Weber.
  % \par}
}

%
% ----- dedication
%
\dedication{\begin{center}to Vitaliy Kuznetsov, Svetlana Kuznetsova, Aleksandr Shishkevich, and Nina Shishkevich who inspired me to be a better student and showed me the impact that professors can have.\end{center}}

%
%

%
% ==========      Text pages
%

\textpages
  
% ========== Chapter 1
 
\chapter {Introduction} \footnote{This thesis expands on the paper by Anna Kuznetsova and Robbie Weber that was accepted to ITiCSE 2026 \cite{paper}.}
Theory courses are foundational for computer scientists---discrete mathematics and probability theory form part of the Mathematical and Statistical Foundation in ACM curricular guidelines \cite{acmcuric}---but the importance of the material is sometimes missed by students. Instructor anecdotes of students wondering why they are learning the material or students feeling unmotivated are common in the literature (e.g., \cite{dmSystematic,remshagen2010making,PhysicsTCS, marketBasketDiscrete,dmPanel,gries2009discrete}), and we, from our own anecdotal experience, had similar concerns.
Ensuring students are motivated is key to students' success. 
Students who are motivated to learn material are more likely to succeed in courses \cite{shell2013associations,lopez2019motivation}. 
Student motivation can even impact retention in the major as a whole \cite{biggers2008student, futureMotivationRetention, peteranetz2018future}.

Aiming to make our foundational theory courses feel more relevant to students, we replaced 3-4 homework problems in both courses with problems focused on applications of the material (rather than simple practice of the material); the problems intentionally mixed applications from future courses and the real world outside of computer science.  

These problems also provided an opportunity to incorporate ethical considerations into these courses, which theory courses often lack.  
We surveyed students on their feelings about relevance of the course at the beginning and end of the term and we collected qualitative feedback from students on the problems. We also ran surveys in terms not using the problems as a control.

In \Cref{ch-2} we discuss related works and highlight our three main contributions. Then in \Cref{ch-3} we describe our course context, the Real World Problem structure, and introduce the 7 Real World Problems. In \Cref{ch-4} we explain our methodology including the survey questions and analysis approach, then in \Cref{ch-5} we present our five main results. In \Cref{ch-6} we discuss limitations of the study and in \Cref{ch-7} we provide best practices for instructors that may want to introduce these or similar Real World Problems to their course. We conclude this thesis with a brief discussion of potential future work.

% ========== Chapter 2
 
\chapter{Related Work} \label{ch-2}
\section{Relevance of Theory Courses}
Adjusting the content of foundational theory courses to better meet interest has been tried in many ways. In discrete math, that has included adding programming \cite{dmProgBT,remshagen2010making,liu2021discrete}, adding computational complexity content \cite{del2021learning, LallAutomata}, adding other applications from future coursework~\cite{applicationsClaim}, and incorporating the same problem across theory and programming courses  \cite{FlatlandModes}. For probability theory, drawing on examples across computer science~\cite{10.1145/1953163.1953245} and adding (programmatic) simulation of experiments \cite{PSSimulation} have been tried.  
Probability has also been suggested as a source of applications in CS1 \cite{PSinIntroPanel}.

While anecdotes of students worrying about relevance are common, actual study of student perception is severely limited.
Course redesign proposals sometimes use end-of-term course evaluations to evaluate student perception, but these are often limited to the topics of the redesign, rather than the relevance of the course as a whole. We are unaware of any other work to survey students directly on their attitudes toward applicability of these courses, though it is sometimes noticed incidentally (e.g., \cite{magda2025students} received qualitative feedback on the topic in discrete math).

We ran a pilot study in our Probability and Statistics course in Spring 2024, and found a surprisingly low fraction of students finding course material to not be relevant, which motivated us to conduct a larger study covering two foundational theory courses. 

\section{Ethics}
Embedding ethics material into technical courses has become more common in recent years, surpassing stand-alone ethics courses \cite{brown2024teaching}.
However, ethics interventions in the literature are far more common in programming and upper-division courses than in theory courses (see Sect. 4.2.4 of Brown et al. \cite{brown2024teaching}).
We are aware of only a few ethics problems used in theory courses. Harvard Embedded EthiCS program deployed a problem in a 11-person upper-division theory course as one of 14 courses studied in \cite{grosz2019embedded}. Some ethics repositories contain a few problems for discrete math \cite{embeddedOrgans, embeddedCore} or upper-division algorithms \cite{embeddedAllocation}, but we are unaware of peer-reviewed versions of these assignments or experience reports with formal evaluation of them. Richard et al. updated combinatorics problems on one homework assignment to be more inclusive, but their focus is on instructor choices impacting students, rather than students learning ethical implications of course material \cite{10.1145/3478431.3499330}.
We are not aware of any papers that provide rich reflection on ethics problems deployed specifically in foundational theory courses.

\section{Contributions} 
Our three main contributions are:
\begin{enumerate}
    \item Our surveys indicate (counter to assumptions in the literature), that most students do believe theory courses are relevant (both before and after the term); these results should motivate further study on student motivation in these courses.  
    \item We provide seven ``Real World Problems'' that can be integrated into discrete math and probability courses that seem to enhance student belief that the courses are relevant to them based on long-form responses. Some of our problems include ethics components, which expand a currently very small library of available ones in intro theory courses. These problems were well-received by students (over $90\%$ recommended reusing them), and can easily be incorporated directly into foundational theory courses.
    \item The structure of the ``Real World Problems'' themselves (see \Cref{sec:problemDesign}), which instructors who want to improve student perceptions of relevance may find adaptable to their courses.  
\end{enumerate}

% ========== Chapter 3
 
\chapter{The Real World Problems} \label{ch-3}
\section{Course Context}
We ran our intervention at the University of Washington; a large, public, R1 university on the West Coast of the United States. The intervention was run across our Discrete Math course (DM) during Autumn quarter 2024 and Probability and Statistics course (PS) during Winter quarter 2025.  
Both courses are required for and restricted to students in the Computer Science and Computer Engineering majors.  They are generally taken in the second- or third-year in the program. DM is a prerequisite for PS. Both are prerequisites for most senior-level courses. Intervention terms had over 300 students enrolled (DM: 324, PS: 354). Control terms were somewhat smaller, but still substantial (DM: 201, PS: 138), occurring in Winter quarter 2025 and Spring quarter 2025, respectively. Weber was the instructor of record for both intervention terms.  

We incorporated three Real World Problems into DM and four Real World Problems into PS, spread across the eight homework assignments for each term.
In addition to the Real World Problems, these courses already include lecture material based on future courses or the real world. For example, DM has a lecture on RSA encryption (connecting to number theory content) and PS uses medical disease testing in Bayes Rule computation examples.

\section{Problem Design} \label{sec:problemDesign}
The Real World Problems were designed to connect a recent course topic to either an application in a future course or to an application in the real world. Some of the applications included CS and non-CS ethical considerations. These problems were not just ``word problems''---standard (fictitious) homework problems phrased in English. Word problems are valuable (the courses already contain them), but the focus of those problems is reinforcing the technical concept. Instead, these Real World Problems focus on the connection between the real world and the course material. Each Real World Problem had most (or all) of these components:
\begin{itemize}
    \item A small technical portion (doing a proof, computing a probability, etc.), which could be graded for correctness.  
    \item An open-ended portion, which asked students to apply concepts to a real world scenario of their choosing (from a provided list or found by the student), which could be graded for completion.
    \item A short written reflection on some aspect of the connection between the real world topic and the course concept, which could be graded for completion.
\end{itemize}

\begin{table*}[h!]
    \begin{tabular}{|p{1.55in}|c|p{2.5in}|c|p{.7in}|}
         \hline \textbf{Topic} & \textbf{Course} & \textbf{Description}  & \textbf{Open} & \textbf{Ethics} \\
          &  &   & \textbf{Ended} &  \\ \hline
         \mbox{Propositional Logic} \mbox{(\Cref{DM-HW2})}& DM & \mbox{Analyze a U.S. Supreme Court case} \mbox{where the interpretation of ``does} \mbox{not have A, B, and C'' is debated:} \mbox{$\neg(a\land b \land c)$ vs. $(\neg a \land \neg b \land \neg c).$}& No & Yes \mbox{(non-CS)} \\ \hline 
         \mbox{Inference Proofs} \mbox{(\Cref{DM-HW4})}& DM & \mbox{Find an example of an incorrect} \mbox{argument (e.g., affirming the} \mbox{consequent) in real life.} & Yes & No\\ \hline 
         \mbox{Induction} \mbox{(\Cref{DM-HW6})}& DM & Perform an induction proof showing correctness of a  piece of code. & No & No\\ \hline
         \mbox{Pigeonhole Principle} \mbox{(\Cref{PS-HW2})}& PS & \mbox{Find and apply the pigeonhole} \mbox{principle to your own scenario.} & Yes & No\\ \hline 
          \mbox{Bayes' Rule} \mbox{(\Cref{PS-HW4})}& PS & \mbox{Critique incorrect arguments based} \mbox{in crime scene DNA testing.} & Yes & Yes \mbox{(non-CS)}\\ \hline 
          \mbox{Random Variable} Modeling \mbox{(\Cref{PS-HW7})}& PS & \mbox{List out (unlisted) assumptions in} \mbox{probability model.} & Yes & No\\ \hline
          \mbox{Machine Learning} Fairness \mbox{(\Cref{PS-HW8})}& PS & \mbox{Interpret and differentiate} \mbox{probabilistic definitions of fairness.} & Yes & Yes \mbox{(in CS)}\\ \hline
    \end{tabular}
    \caption{Topics for Real World Problems in DM and PS.}
    \label{tab:DM-problems}
\end{table*}

Details of all the Real World Problems are shown in \Cref{tab:DM-problems} \footnote{Updated problems with commentary for instructors are available in Appendix A - Appendix G and on \href{https://github.com/annak4397/UW-CSE-Real-World-Problem-Research-Project/tree/main/problems/updated-problems}{GitHub}. Exact problems that were given to students during intervention quarters can be found in this folder on \href{https://github.com/annak4397/UW-CSE-Real-World-Problem-Research-Project/tree/main/problems/original-problems}{GitHub}.}. The goal of these problems was to show students the relevance of the material, while still reinforcing the concepts. By leaving many of them open-ended, we hoped students would find topics they were excited about, and thus connect more with the material. 
% ========== Chapter 4
 
\chapter{Methods} \label{ch-4}
Students were given online surveys through Gradescope at the beginning (week 1 or 2, called ``pre-survey'') and end of term (week 10, due between last class period and final exam, called ``post-survey''). Our Institutional Review Board examined our proposed intervention and declared it exempt from further review as being standard educational practice.

\section{Survey Questions}

\paragraph{Relevance}
These questions were given with a Likert Scale (Strongly Disagree, Disagree, Neutral, Agree, Strongly Agree) and asked in both surveys. 

\textbf{Q1} I believe that theoretical computer science is relevant to real life. 

\textbf{Q2} I believe that theoretical computer science is relevant to my future career as a 

computer scientist. 

\textbf{Q3} I believe that [discrete math/probability and statistics] is relevant to real life.

\textbf{Q4} I believe that [discrete math/probability and statistics] to my future career as a 

computer scientist.

\paragraph{Problem Feedback} These questions were asked for each Real World Problem during intervention quarters in the post-survey.

\textbf{Q5} Please rate how you felt about this problem. (Strongly Disliked, Disliked, Neutral, 

Liked, Strongly Liked)

\textbf{Q6} Do you have any thoughts that you would like to share specifically about this problem?

(textbox)

\paragraph{General Feedback} These questions were during intervention quarters in the post-survey.

\textbf{Q7} How did the Real World problems affect your understanding of the concepts learned 

in class? (textbox)

\textbf{Q8} How did the Real World problems affect your view of the real world? (textbox)

\textbf{Q9} Should Real World problems be included in future quarters? (Yes/No multiple 

choice); please explain why. (textbox)

\textbf{Q10} Do you have any other thoughts on the real world problems that you would like to 

share with us? (textbox)

\section{Methodology}
In intervention terms (24 Autumn DM; 25 Winter PS), students received the Relevance Questions (Q1-Q4) and a textbox for ``any other comments'' in the pre-survey and all questions (Q1-Q10) in the post-survey. In control terms (25 Winter DM, 25 Spring PS), students received the Relevance Questions and a textbox for ``any other comments'' in both the pre- and post-survey. 
 Control quarters were taught by other instructors without the Real World Problems, but with similar lecture content to intervention quarters. Students who were retaking DM or PS were excluded from numerical data, as they may have seen pilot versions of the Real World Problems.

 Both courses contained ``concept checks'': a short quiz due at the start of each lecture. A small part of students' course grade required achieving $80\%$ or higher correctness on these quizzes. To incentivize participation, submitted surveys were counted as a bonus concept check. In intervention terms, every homework assignment had a 1-point feedback problem; homework assignments with a Real World Problem contained an additional optional question for feedback on that problem (phrased like Q6). Responses there and to Q6 on the post-survey were combined per student and analyzed together.
 
\section{Analysis}
All survey questions were optional. Some students skipped questions, leading to slightly different response totals by question. We show the total number of responses as $n$ in all applicable tables. For Likert Scale questions, students were shown qualitative descriptions of categories; for tables we convert these responses to a 1-5 numerical scale (SD=1, SA/SL=5). 
 In pre-surveys (only), we added an additional direction that if students were unsure what terms (e.g., ``discrete math'') meant, they should mark ``neutral.''

Long-form responses were grouped by question for coding. Codes were found inductively from reading student responses. In the first phase, both authors read the first 50 responses to each question, applied codes independently, and suggested updates to the codebook. Disagreements were resolved and the codebook was updated by discussion. In the second phase, the remaining responses were coded by an author, with the other checking the coding and marking disagreements. All disagreements were again resolved by discussion. 

% ========== Chapter 5

\chapter{Results} \label{ch-5}
From statistical analysis of survey questions and coding of qualitative feedback, we observed:
\begin{itemize} 
    \item Students found the courses to be relevant. (\Cref{sec:result1})
    \item That feeling of relevance was stronger with the Real World Problems. (\Cref{sec:result2})
    \item Students liked having the Real World Problems. (\Cref{sec:result3})
    \item Real World Problems had a positive impact on the way students learned in the course. (\Cref{sec:result4})
    \item Students were able to identify the ethical implications surfaced by the Real World Problems. (\Cref{sec:result5})
\end{itemize} 

In the rest of the paper, we report the number of codes applied across responses to Q5-Q10 in the intervention post-surveys, broken down by course in parentheses. In total, 3131 codes were applied. Codes explicitly mentioned in the paper are shown in \Cref{tab:codesPaper} and the full code book with example quotes is included in \Cref{Codebook}.

\begin{longtable}{|p{4.75in}|c|c|}
    \hline
       \textbf{Code}  & $n$ \textbf{in DM} & $n$ \textbf{in PS}\\ \hline 
       Saw course concepts in the real world. & 336 & 319 \\ \hline 
       Saw course concepts in future courses and/or careers. & 40 & 9 \\ \hline 
       Viewed course content as important or useful. & 31 & 78 \\ \hline \hline
       Positive comment about Real World Problem(s). & 302 & 361 \\ \hline 
       Negative comment about Real World Problem(s). & 40 & 48 \\ \hline 
       Neutral comment about Real World Problem(s).& 17 & 34 \\ \hline
       Inconsistent comments about Real World Problem(s). & 5 & 8 \\ \hline \hline 
       Real World Problem(s) were interesting or engaging & 136 & 73 \\ \hline 
       Real World Problem(s) caused students to have a small change in perspective about the real world & 52 & 5 \\ \hline 
       Real World Problem(s) caused students to think about the course concepts differently. & 23 & 18 \\ \hline 
       Appreciated having a different kind of problem.  & 19 & 14 \\ \hline 
       Real World Problem(s) made it clear why they were learning the content. & 18 & 19 \\ \hline 
       Asked for more Real World Problems to be added to homework. & 15 & 4 \\ \hline \hline
       Real World Problem(s) helped them understand the course concepts better. & 108 & 134 \\ \hline 
       Real World Problem(s) did not have an effect on their learning. & 18 & 31 \\ \hline 
       Real World Problem(s) had a negative impact on their learning. & 2 & 0 \\ \hline \hline
       Mentioned confusion with regards to Real World Problem(s). & 50 & 43 \\ \hline 
       Struggled to find a real world example for Real World Problem(s). & 21 & 46 \\ \hline 
       Spent significant amount of time on the Real World Problem(s) & 11 & 50 \\ \hline 
       Worried about getting open-ended questions of Real World Problem(s) right. & 11 & 15 \\ \hline
       Found the amount of reading with Real World Problem(s) off-putting. & 8 & 2 \\ \hline \hline
       Found Real World Problem(s) challenging. & 28 & 14 \\ \hline 
       Found Real World Problem(s) easier. & 7 & 18 \\ \hline \hline   
       Liked the open-ended portion of Real World Problem(s). & 16 & 47 \\ \hline 
       Disliked the open-ended portion of Real World Problem(s). & 2 & 8 \\ \hline  
    \caption{Codes from qualitative feedback mentioned in the paper.}
    \label{tab:codesPaper}
\end{longtable}

\section{Students found DM and PS relevant} \label{sec:result1}
\begin{figure}[htbp]
\begin{subfigure}[t]{\textwidth}
\vspace{-25pt}
    \begin{subfigure}{0.5\textwidth}
        \centering
        \includegraphics[width=1\linewidth]{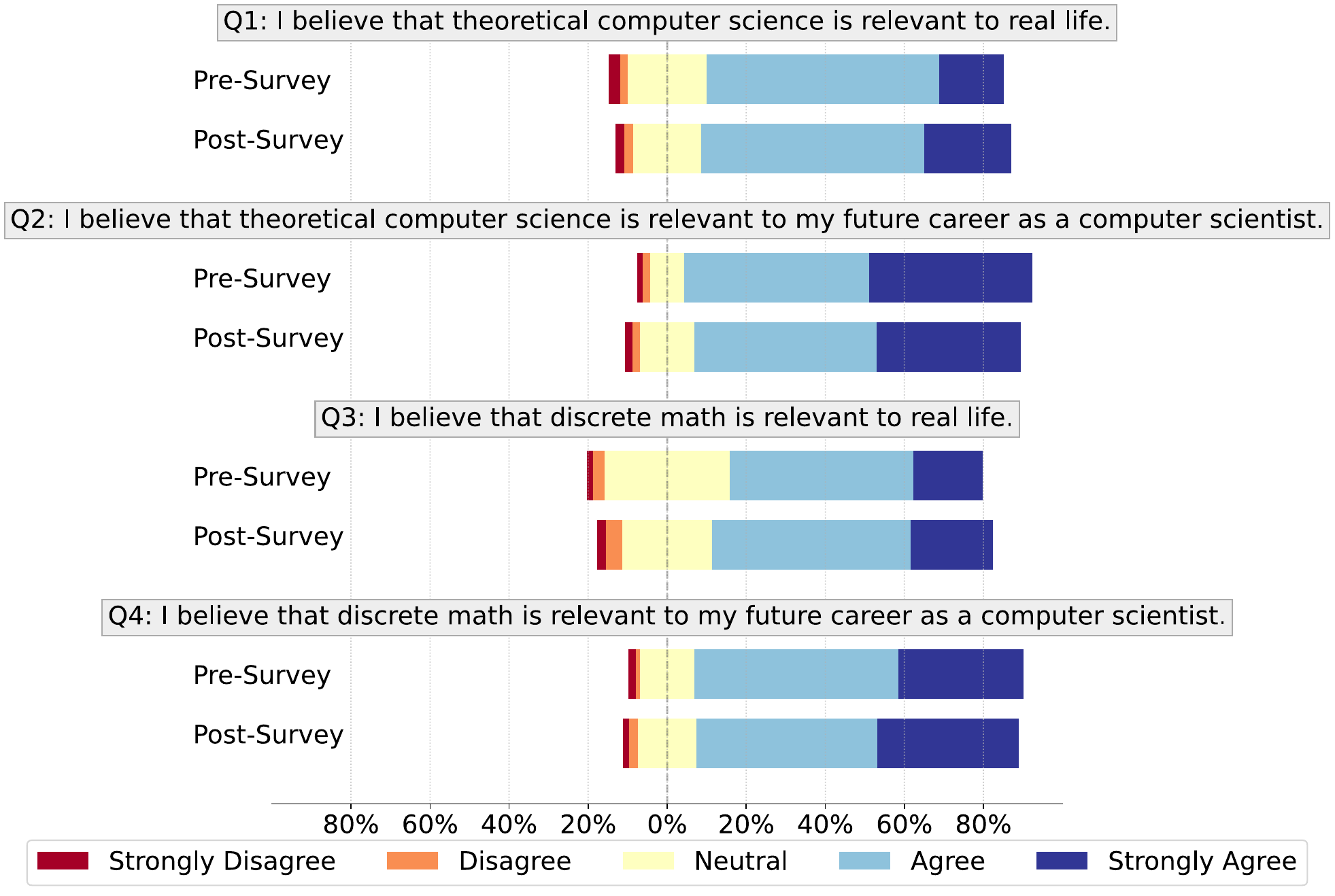}
        \caption{DM Intervention Term}
    \end{subfigure}
    \begin{subfigure}{0.5\textwidth}
        \centering
        \includegraphics[width=1\linewidth]{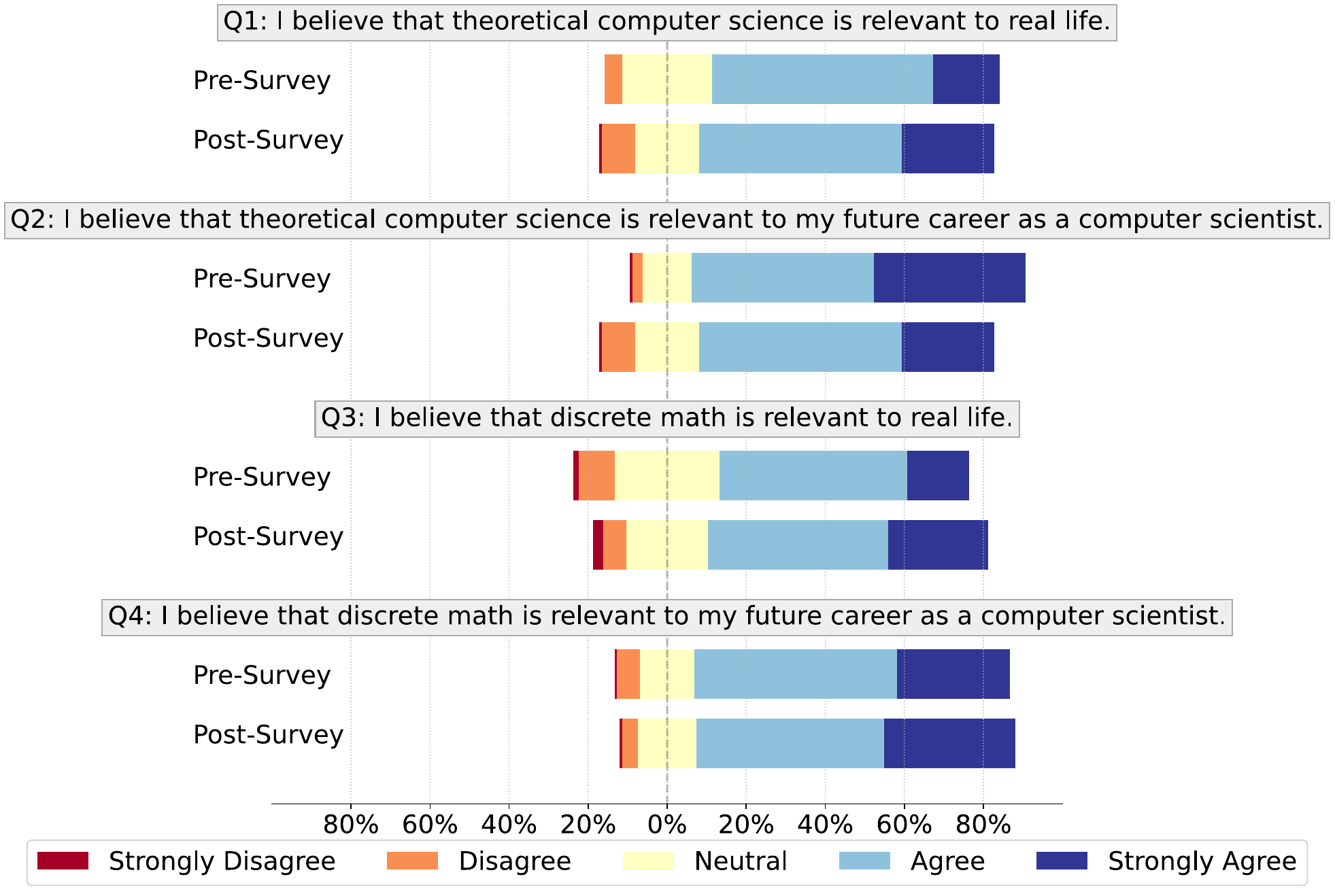}
        \caption{DM Control Term}
        
    \end{subfigure}
\label{fig:LikertDM}
\end{subfigure}
\begin{subfigure}[t]{\textwidth}
\vspace{-10pt}
   \begin{subfigure}{0.5\textwidth}
        \centering
        \includegraphics[width=1\linewidth]{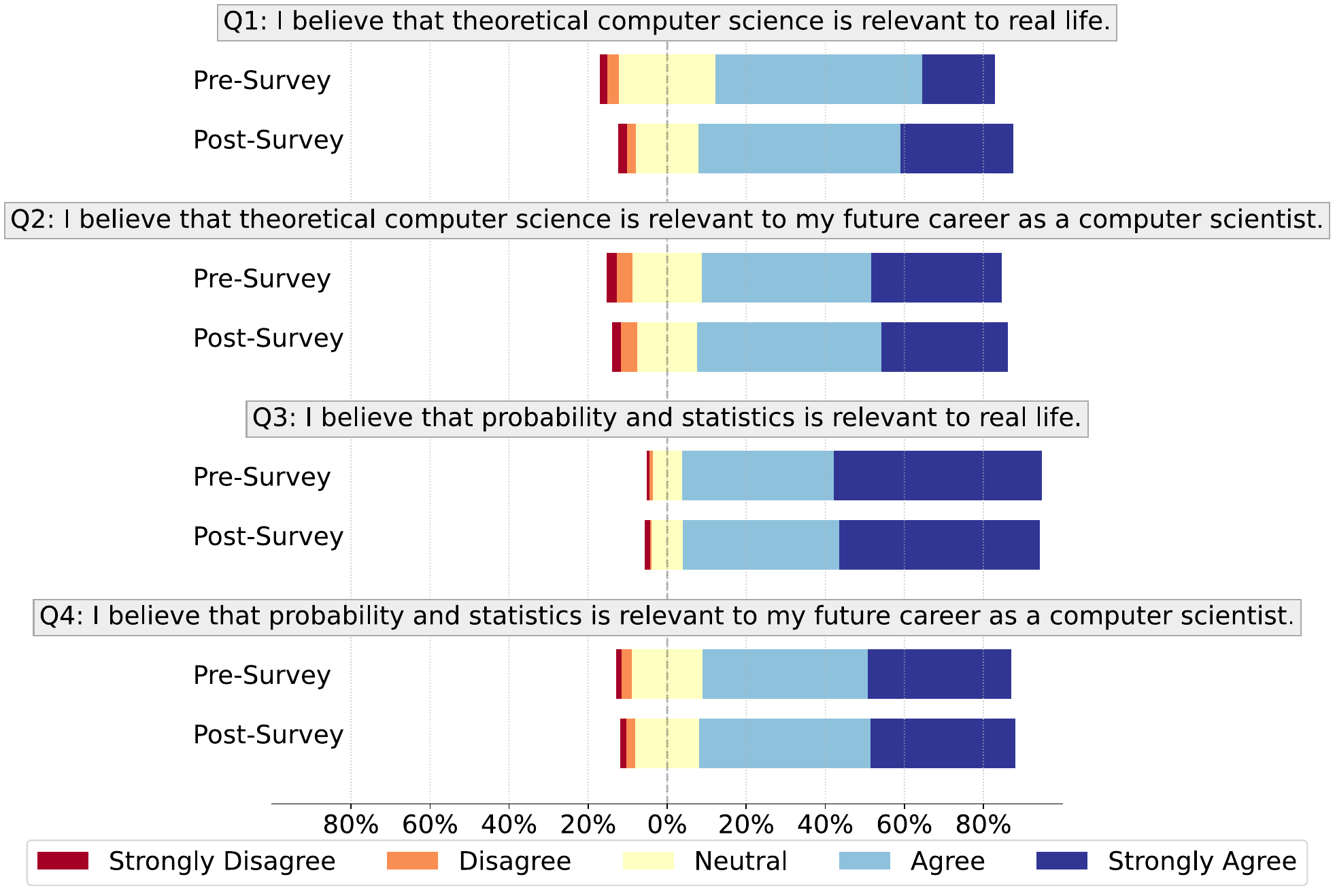}
        \caption{PS Intervention Term}
    \end{subfigure}
    \begin{subfigure}{0.5\textwidth}
        \centering
        \includegraphics[width=1\linewidth]{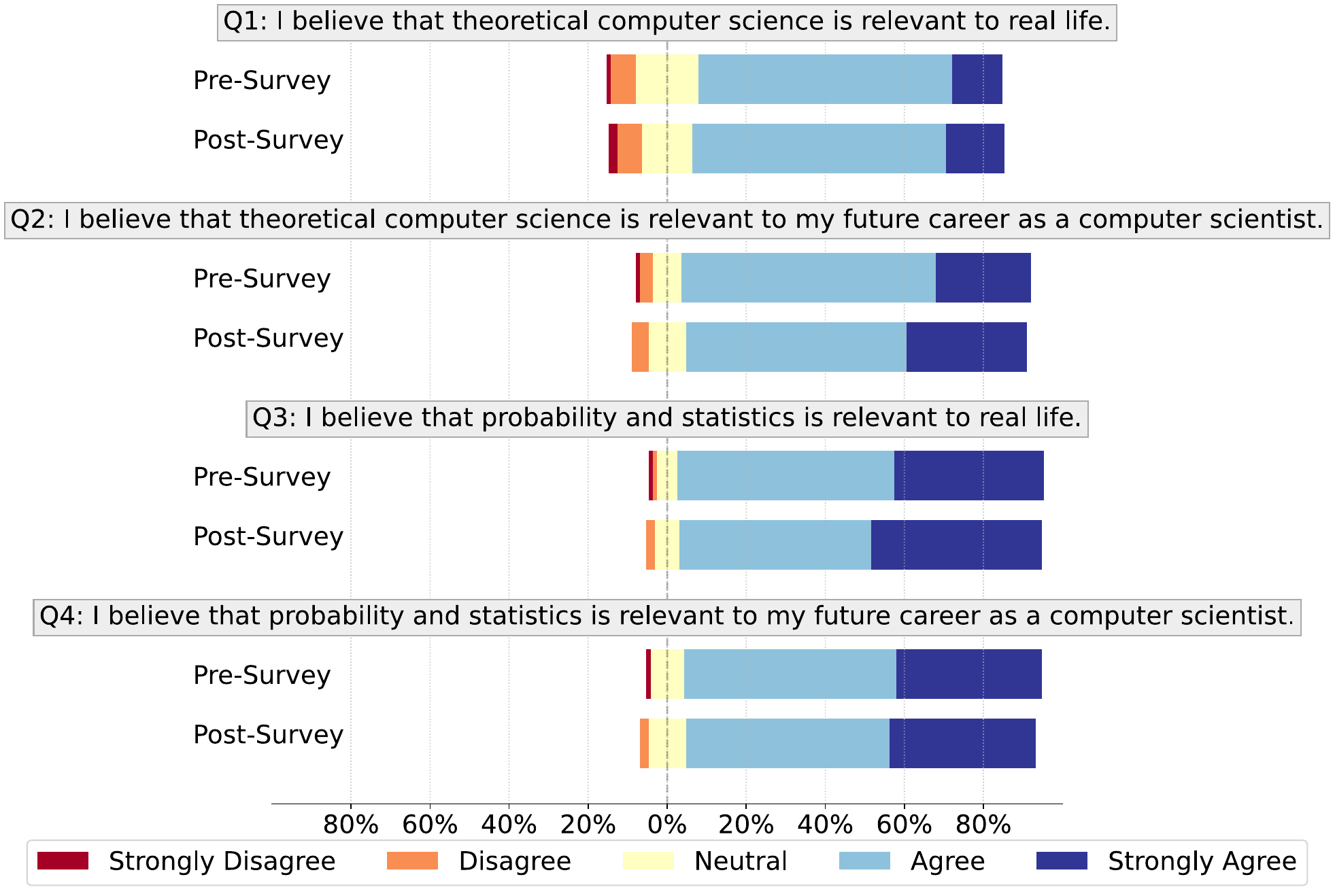}
        \caption{PS Control Term}
    \end{subfigure}
\label{fig:LikertPS}
\end{subfigure}
\begin{subfigure}[t]{\textwidth}
\vspace{-10pt}
   \begin{subfigure}{0.5\textwidth}
        \centering
        \includegraphics[width=1\linewidth]{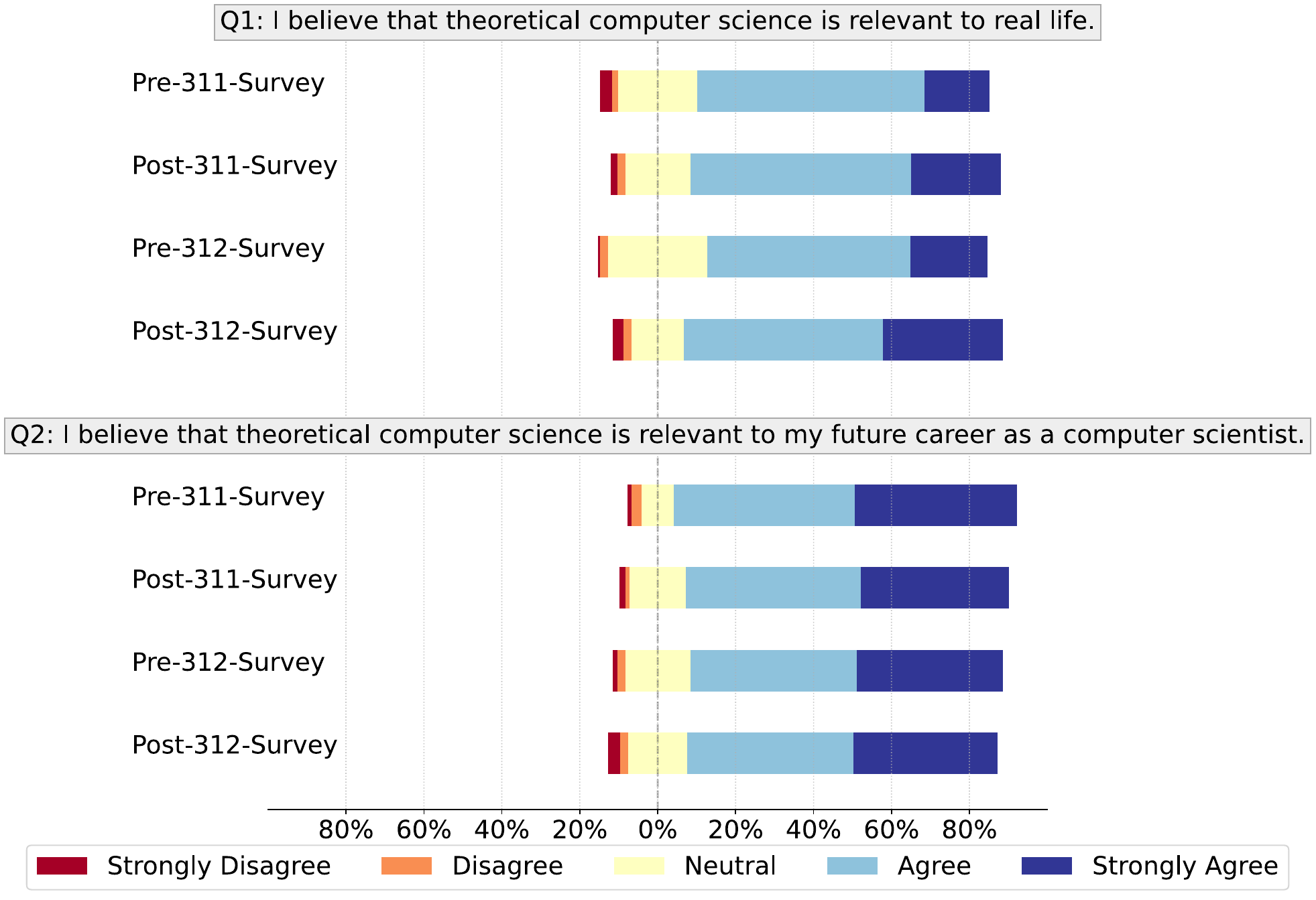}
        \caption{Intervention Terms}
    \end{subfigure}
    \begin{subfigure}{0.5\textwidth}
        \centering
        \includegraphics[width=1\linewidth]{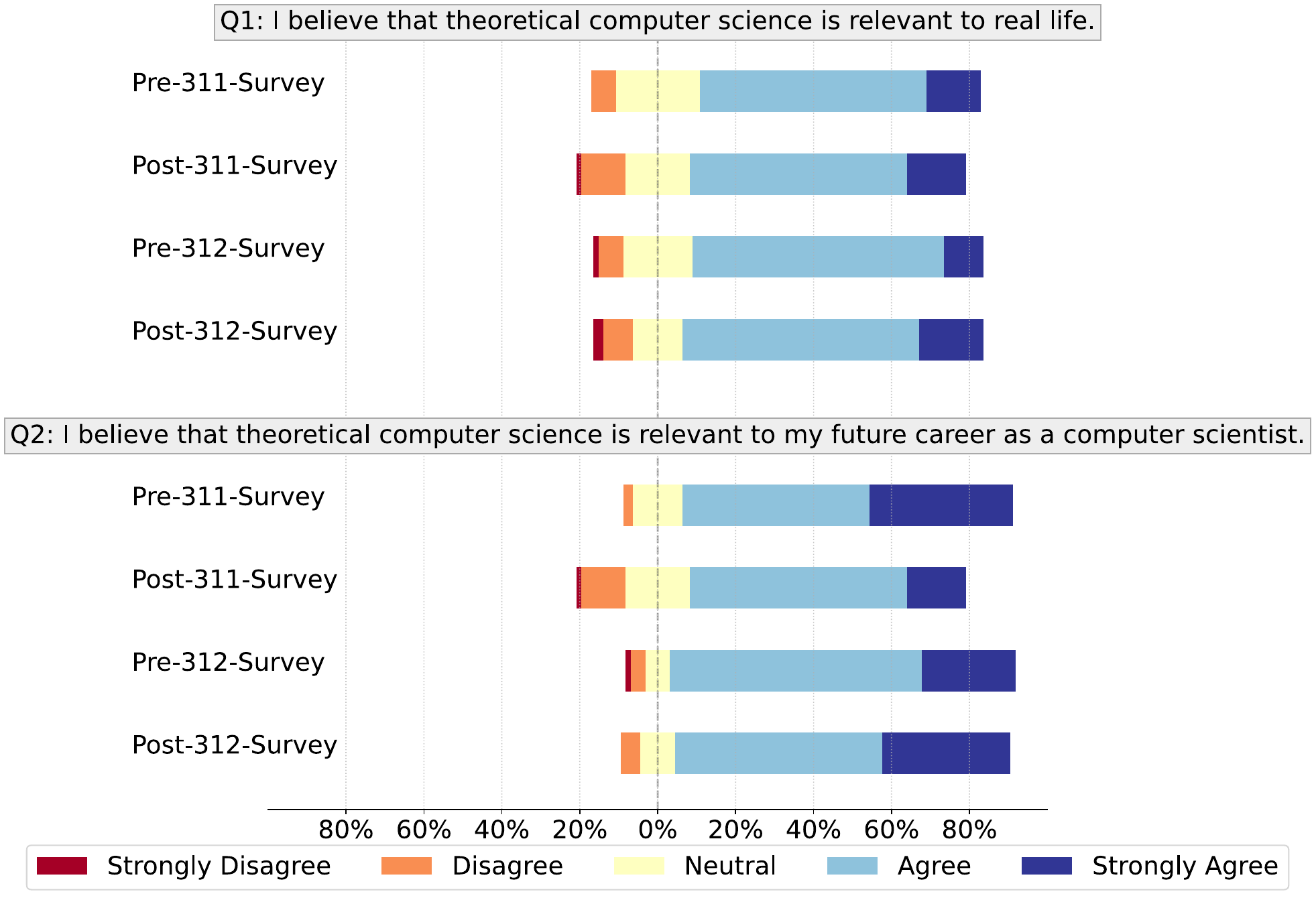}
        \caption{Control Terms}
    \end{subfigure}
\label{fig:LikertBoth}
\end{subfigure}
\vspace{-15pt}
\caption{Student responses to Likert Scale questions. The bars show Strongly Disagree (left) to Strongly Agree (right); centered at midpoint of Neutral.}
\label{fig:Likert}
\end{figure}

Despite getting individual concerns about applicability from students in the past, the majority of students in all surveys (pre- and post-surveys, in all control and intervention quarters) found theoretical computer science and the specific courses applicable to both real life and their future careers. Responses to Q1-Q4 for the intervention and control terms in DM, PS, and in the sequence of courses are visualized in \Cref{fig:LikertBoth}.

Student feedback also showed that students found the courses relevant. Students mentioned realizing (or already knowing) that the content was important or useful (DM: 32, PS:78), that they saw course concepts in the real world (DM: 336, PS: 319) or future courses and/or careers (DM: 40, PS: 9), that problems caused students to think about course concepts differently (DM: 23, 18), that problems caused students to view the real world differently (DM: 52, PS: 5), or made it clear why they were learning the content (DM: 18, PS: 19). One student said the Real World Problems ``made me reflect on `does any of this matter in real life' I'm already starting to see combinatorics and probabilities everywhere!'' Another student said the problems ``Makes me think a lot more about the applicability of the theory we're learning in the [DM], [PS] sequence. Makes me feel more optimistic about the CS degree that I'm pursuing, to become a better thinker and problem solver.''

\section{Impact of Real World Problems} \label{sec:result2}
\subsection{Analysis of Full Sequence}
We hoped to evaluate whether the problems affected students' view of the relevance of the courses. As responses started so high, there was relatively little room for improvement, but some can still be seen. We perform an analysis of Q1 (theoretical CS relevant to real life) and Q2 (theoretical CS relevant to career) on students who took the full sequence of courses. This means they either took both intervention quarters or both control quarters. To be included in the analysis they needed to complete both Q1 and Q2 on both the pre-survey in DM and the post-survey in PS. Students who were retaking either course were excluded. The resulting groups represent more than half of the students in the study. Student responses for Q1 and Q2 are in \Cref{tab:Sequence}. 

\begin{table}
    \centering
    \begin{tabular}{|c|c|c|c||c|c|c|c|c|}
    \hline
       \textbf{Group}  & \textbf{Qn} & $n$ & \textbf{Avg.} & \textbf{SD} & \textbf{D} & \textbf{N} & \textbf{A} & \textbf{SA}  \\ \hline
         Intervention Pre DM& Q1 & 192 & 3.84 & 6 & 3 & 39 & 112 & 32\\ \hline
         Intervention Post PS& Q1 & 192 & 4.05 & 5 & 4 & 26 & 98 & 59\\ \hline
         Intervention Pre DM & Q2 & 192 & 4.25& 2 & 5 & 16 & 89 & 80\\ \hline
         Intervention Post PS & Q2 & 192 & 4.08 & 6 & 4 & 29 & 82 & 71\\ \hline \hline
         Control Pre DM& Q1 & 79 & 3.80 & 0 & 5 & 17 & 46 & 11\\ \hline
         Control Post PS& Q1 & 79 & 3.81 & 2 & 6 & 10 & 48 & 13\\ \hline
         Control Pre DM & Q2 & 79 & 4.19& 0 & 2 & 10 & 38 & 29\\ \hline
         Control Post PS & Q2 & 79 & 4.14 & 0 & 4 & 7 & 42 & 26\\ \hline
    \end{tabular}
    \caption{Results for  Q1 (theoretical CS relevant to real life) and Q2 (theoretical CS relevant to career)  for students taking both intervention or both control quarters.}
    \label{tab:Sequence}
\end{table}

 Chi-Square ($\chi^2$) tests of significance were run to evaluate the impact of these problems on student responses. Due to small numbers of ``strongly disagree'' responses, all tests were run with ``disagree'' and ``strongly disagree'' combined into one entry. A rule of thumb for $\chi^2$ tests is to avoid cells with expected values below 5, though the exact recommendations vary; see \cite{kroonenberg2018tale} for various versions of the rule. We combined D and SD for \textbf{all} tests for consistency among them. Results of $\chi^2$ tests are shown in \Cref{tab:SequenceChiSq}. 

\begin{table}
    \centering
    \begin{tabular}{|c|c|c|c|c|}
    \hline
    \textbf{Compared Groups}    & \textbf{Qn} & $\chi^2$  & $p$ & \textbf{Sig}  \\ \hline
    Intervention Pre vs. Control Pre & Q1 & 0.5968 & .8972 & \\ \hline
    Intervention Pre vs. Control Pre & Q2 & 1.6789 & .6416 &  \\ \hline \hline 
     Intervention Post vs. Control Post & Q1 & 7.9455 & .0472 & * \\ \hline
    Intervention Post vs. Control Post & Q2 & 3.2408 & .3560 & \\ \hline  \hline 
    Intervention Pre vs. Intervention Post & Q1 & 11.5443 & .0091 & ** \\ \hline
    Intervention Pre vs. Intervention Post & Q2 & 5.1079 & .1641 &  \\ \hline \hline 
    Control Pre vs. Control Post & Q1 & 2.7163 & .4375 & \\ \hline 
    Control Pre vs. Control Post & Q2 & 1.5597 & .6686 & \\ \hline 
    
    \end{tabular}
    \caption{$\chi^2$-test statistics and $p$-values comparing intervention and control sequence groups on Q1 and Q2. Pre represents before DM; post represents after PS. $p$-values below .05 are marked with *; $p$-values below .01 are marked with **.}
    \label{tab:SequenceChiSq}
\end{table}

From the results, we see that before taking DM, the control and intervention distributions are similar (see \Cref{tab:SequenceChiSq}, Rows 1,2), but the intervention does seem to have a noticeable impact on responses to Q1 in the post-survey. Specifically, the intervention group has fewer neutral (from 39 to 26) and many more strongly agree responses (from 32 to 59) after taking the courses, while the control replies are only minorly changed after taking the sequence. That is, the problems seem to give students a stronger feeling of relevance to real life than they had without them. 

Conversely, the effect on Q2 (theoretical CS relevant to career) is much less clear. Absolute changes among entries are smaller than for Q1, and tend toward neutral in the intervention group; no entries in the control group changed by more than 4 students of 79. No $\chi^2$ test was significant for Q2 among these groups (See \Cref{tab:SequenceChiSq}, Rows 2,4,6,8). 
We conjecture that our focus on a variety of applications (only one of the problems directly mentions code) may have lessened the impact on Q2. 

The full sequence of both courses seems to strengthen students' belief that theory courses will be relevant to their real life relative to control. Regardless of including the problems or not, students enter and leave the course with a belief that theory courses will be relevant to their careers, though these opinions seem to slightly decrease in strength across the sequence.  

\subsection{Analysis of Individual Courses}
As the courses do not have to be taken in consecutive quarters, examining single quarter effects gave larger datasets. Results here seem somewhat less clear than for the full sequence (perhaps because they represent just one term of intervention rather than two). Survey responses in \Cref{tab:311OnlyIntervention} and \Cref{tab:311OnlyControl} show general agreement on relevance of the material in both control and intervention quarters before and after the course. At most $11$\% disagreed (or strongly disagreed) and at least $62$\% agreed (or strongly agreed). $\chi^2$-tests on DM show an effect on Q2 (instead of Q1, as in the full sequence); see \Cref{tab:311ChiSq}. 

Results for DM suggest that not having Real World Problems resulted in students believing that Theoretical Computer Science is less relevant to their future careers than they believed at the start of DM. As shown in \Cref{tab:311ChiSq} there was no statistical significance in responses for the pre-survey questions between the intervention and control quarters (rows 1-4), but there was statistical significance when comparing responses to Q2 (row 6) between the intervention and control at the end of the quarter. The most notable difference comes from a decrease in students Strongly Agreeing to Q2 in the control quarter (see row 14 in \Cref{tab:311ChiSq} and rows 3 and 4 in  \Cref{tab:311OnlyControl}). The Real World problems seem to be more effective in DM at maintaining belief in career relevance.   

\begin{table}
    \centering
    \begin{tabular}{|c|c|c|c||c|c|c|c|c|}
    \hline
       \textbf{Group}  &\textbf{Qn}& $n$ & \textbf{Avg.} & \textbf{SD} & \textbf{D} & \textbf{N} & \textbf{A} & \textbf{SA}  \\ \hline
         Intervention Pre & Q1 & 269 & 3.84& 8 & 5 & 54 & 158 & 44\\ \hline
         Intervention Post & Q1 & 269 & 3.94 & 6 & 6 & 46 & 152 & 59\\ \hline
         Intervention Pre & Q2 & 269 & 4.25& 4 &5  & 23 & 126 & 111\\ \hline
         Intervention Post & Q2 & 269 & 4.13 & 5 & 5 & 37 & 124& 98\\ \hline
         Intervention Pre & Q3 & 269 & 3.75 &4 & 8 & 85 & 125 & 47\\ \hline
         Intervention Post & Q3 & 269 & 3.83 & 6 & 11  & 61 & 135&56 \\ \hline
         Intervention Pre & Q4 & 269 & 4.10 & 5 & 3 & 37 & 139 & 85\\ \hline
         Intervention Post & Q4 & 269 & 4.12 & 4 & 6 & 40 & 123& 96\\ \hline
    \end{tabular}
    \caption{Responses for students taking DM intervention}
    \label{tab:311OnlyIntervention}

    \begin{tabular}{|c|c|c|c||c|c|c|c|c|}
    \hline\textbf{Group}  &\textbf{Qn}& $n$ & \textbf{Avg.} & \textbf{SD} & \textbf{D} & \textbf{N} & \textbf{A} & \textbf{SA}  \\ \hline
         Control Pre & Q1 & 154 & 3.85 & 0 & 7 & 35 & 86 & 26\\ \hline
         Control Post & Q1 & 154 & 3.88 & 1 & 13 & 25 & 79 & 36\\ \hline
         Control Pre & Q2 & 154 & 4.19& 1 & 4 &  19 & 71 & 59\\ \hline
         Control Post & Q2 & 154 & 3.88 & 1 & 13 & 25 & 79 & 36\\ \hline
         Control Pre & Q3 & 154 & 3.67 & 2 & 14 & 41 & 73 & 24\\ \hline
         Control Post & Q3 & 154 & 3.85 &  4 & 9 & 32 & 70 & 39\\ \hline
         Control Pre & Q4 & 154 & 4.01 & 1  & 9 & 21 & 79 & 44\\ \hline
         Control Post & Q4 & 154 & 4.08  &  1&  6& 23 & 73 & 51\\ \hline
    \end{tabular}
    \caption{Responses for students taking DM control}
    \label{tab:311OnlyControl}
\end{table}

\begin{table}
    \centering
    \begin{tabular}{|c|c|c|c|c|}
    \hline
    \textbf{Compared Groups}    & \textbf{Qn} & $\chi^2$  & $p$ & \textbf{Sig}  \\ \hline
    Intervention Pre vs. Control Pre & Q1 & 7.074 & 0.069576 & \\ \hline
    Intervention Pre vs. Control Pre & Q2 & 1.6416 & 0.649999 & \\ \hline
    Intervention Pre vs. Control Pre & Q3 & 6.2402 & 0.10049 & \\ \hline
    Intervention Pre vs. Control Pre & Q4 & 3.1487 &  0.369256 &  \\ \hline \hline 
    Intervention Post vs. Control Post & Q1 & 4.0364 & 0.257565 &  \\ \hline
    Intervention Post vs. Control Post & Q2 & 11.2154 &  0.010617 &  * \\ \hline
    Intervention Post vs. Control Post & Q3 & 2.1201 & 0.54785 &  \\ \hline
    Intervention Post vs. Control Post & Q4 & 0.4131 & 0.937527 & \\ \hline  \hline 
    Intervention Pre vs. Intervention Post & Q1 & 2.9806 & 0.394627 &  \\ \hline
    Intervention Pre vs. Intervention Post & Q2 & 4.1439 &  0.246338 &  \\ \hline
    Intervention Pre vs. Intervention Post & Q3 & 5.9783 &  0.112671 &  \\ \hline
    Intervention Pre vs. Intervention Post & Q4 & 1.9847 & 0.575586 &  \\ \hline \hline 
    Control Pre vs. Control Post & Q1 & 5.9099 & 0.116078 & \\ \hline 
    Control Pre vs. Control Post & Q2 & 11.0764 & 0.01132 & * \\ \hline 
    Control Pre vs. Control Post & Q3 & 5.0543 & 0.167864 & \\ \hline 
    Control Pre vs. Control Post & Q4 & 1.373 & 0.711887 & \\ \hline 
    \end{tabular}
    \caption{$\chi^2$-test statistics and $p$-values comparing intervention and control DM groups on Q1-Q4. Pre represents before DM; post represents after DM. $p$-values below .05 are marked with *; $p$-values below .01 are marked with **.}
    \label{tab:311ChiSq}
\end{table}

Results for PS were surprising because it appears that the groups were different before the start of the quarter. We found statistical significance in responses to the pre-survey in three out of the 4 questions (rows 2-4 in \Cref{tab:312ChiSqNew}). Due to this significance we do not believe any statistical significance in post-survey responses are due to the Real World Problems. Instead, we conjecture that students having done the Real World problems in DM in 24Au, waiting a term, and taking PS in 25Sp are having a measurable impact (though this may also be a cohort effect or other source, see \Cref{ch-6}). It is worth pointing out that there is statistical significance in the Intervention Pre vs. Post responses to Q1 (row 9 in \Cref{tab:312ChiSqNew}). We believe that the Real World Problems were the reason for 27 more students Strongly Agreeing with Q1 at the end of the intervention quarter (see rows 1-2 in \Cref{tab:312OnlyIntervention}).

\begin{table}
    \centering
    \begin{tabular}{|c|c|c|c||c|c|c|c|c|}
    \hline
       \textbf{Group}  &\textbf{Qn}& $n$ & \textbf{Avg.} & \textbf{SD} & \textbf{D} & \textbf{N} & \textbf{A} & \textbf{SA} \\ \hline
         Intervention Pre & Q1 & 270 & 3.83 & 5 & 8 & 66 & 141 & 50 \\ \hline
         Intervention Post & Q1 & 270 & 4.01 & 6 & 6 & 43 & 138 & 77\\ \hline
         Intervention Pre & Q2 & 270 & 4.00 & 7 &  11& 47 & 116 & 89\\ \hline
         Intervention Post & Q2 & 270 & 4.02  & 6 & 11 & 41 & 126 & 86\\ \hline
         Intervention Pre & Q3 & 270 & 4.41 &2  & 2 & 20 & 104 & 142 \\ \hline
         Intervention Post & Q3 & 270 & 4.38 & 4 & 1 & 21 & 107 & 137\\ \hline
         Intervention Pre & Q4 & 270 & 4.09 & 4 & 7 & 48 & 113 & 98\\ \hline
         Intervention Post & Q4 &270 &  4.11 & 4 & 6 & 44 & 117 & 99\\ \hline
    \end{tabular}
    \caption{Responses for students taking PS intervention}
    \label{tab:312OnlyIntervention}

    \begin{tabular}{|c|c|c|c||c|c|c|c|c|}
    \hline
       \textbf{Group}  &\textbf{Qn}& $n$ & \textbf{Avg.} & \textbf{SD} & \textbf{D} & \textbf{N} & \textbf{A} & \textbf{SA} \\ \hline
         Control Pre & Q1 & 95 & 3.81 & 1 & 6 & 15 & 61 & 12 \\ \hline
         Control Post & Q1 & 95 & 3.83 & 2 & 6 & 12 & 61 & 14 \\ \hline
         Control Pre & Q2 & 95 & 4.07 & 1 & 3 & 7 & 61 & 23 \\ \hline
         Control Post & Q2 & 95 & 4.13 & 0 & 4 & 9 & 53 & 29\\ \hline
         Control Pre & Q3 & 95 & 4.27 & 1 & 1  & 5 &  52 & 36\\ \hline
         Control Post & Q3 & 95 & 4.33 & 0 & 2 & 6 & 46 & 41 \\ \hline
         Control Pre & Q4 & 95 & 4.25 & 1 & 0 & 8 & 51 & 35\\ \hline
         Control Post & Q4 & 95 & 4.23 & 0 & 2 & 9 & 49 & 35 \\ \hline
    \end{tabular}
    \caption{Responses for students taking PS control}
    \label{tab:312OnlyControl}
\end{table}

\begin{table}
    \centering
    \begin{tabular}{|c|c|c|c|c|}
    \hline
    \textbf{Compared Groups}    & \textbf{Qn} & $\chi^2$  & $p$ & \textbf{Sig}  \\ \hline
    Intervention Pre vs. Control Pre & Q1 & 6.4671 & 0.090968 &  \\ \hline
    Intervention Pre vs. Control Pre & Q2 & 13.7872 & 0.00321 & ** \\ \hline
    Intervention Pre vs. Control Pre & Q3 & 8.0759 & 0.044468 & * \\ \hline
    Intervention Pre vs. Control Pre & Q4 & 8.1568 & 0.042879 & * \\ \hline \hline 
    Intervention Post vs. Control Post & Q1 & 10.0996 & 0.017738 & * \\ \hline
    Intervention Post vs. Control Post & Q2 & 3.4366 & 0.329075 &  \\ \hline
    Intervention Post vs. Control Post & Q3 & 2.3509 & 0.502842 &  \\ \hline
    Intervention Post vs. Control Post & Q4 & 3.85 & 0.278119 &  \\ \hline  \hline 
    Intervention Pre vs. Intervention Post & Q1 & 10.6656 & 0.013678 & * \\ \hline
    Intervention Pre vs. Intervention Post & Q2 & 0.9023 & 0.824869 &  \\ \hline
    Intervention Pre vs. Intervention Post & Q3 & 0.2678 & 0.965973 &  \\ \hline
    Intervention Pre vs. Intervention Post & Q4 & 0.2962 & 0.960747 &  \\ \hline \hline 
    Control Pre vs. Control Post & Q1 & 0.5538 & 0.906912 &  \\ \hline 
    Control Pre vs. Control Post & Q2 & 1.5037 & 0.681414 &  \\ \hline 
    Control Pre vs. Control Post & Q3 & 0.7829 & 0.853545 &  \\ \hline 
    Control Pre vs. Control Post & Q4 & 0.4322 & 0.933522 &  \\ \hline 
    \end{tabular}
    \caption{$\chi^2$-test statistics and $p$-values comparing intervention and control PS groups on Q1-Q4. Pre represents before PS; post represents after PS. $p$-values below .05 are marked with *; $p$-values below .01 are marked with **.}
    \label{tab:312ChiSqNew}
\end{table}
\section{Students liked real world problems} \label{sec:result3}
Q5 asked students at the end of each intervention term to rate each problem on a 5 point scale 
%(Strongly Disliked, Disliked, Neutral, Liked, Strongly Liked) 
and Q9 asked whether we should continue the questions. Students reported liking the problems. All had a majority reporting liking (or strongly liking) the problem ($56\%$ or more) and less than $11\%$ reporting disliking (or strongly disliking) the problem. More than $90\%$ of students selected yes on Q9: 254/265 in DM and 268/294 in PS.

\begin{table}[]
    \centering
    \begin{tabular}{|c|c|c||c|c|c|c|c|}
    \hline
      \textbf{Topic} & $n$ & \textbf{Avg.} & \textbf{SD} & \textbf{D} & \textbf{N} & \textbf{L} & \textbf{SL} \\ \hline 
Propositional Logic  & 276 &3.82 & 3	& 15	& 64	& 140	& 54		\\ \hline 
Inference Proofs  &  274 & 3.61 & 5	& 19	& 89	& 125	& 36		\\ \hline 
Induction  & 273 & 3.79 & 6&11	& 77	& 119	& 60 \\ \hline \hline
Pigeonhole Principle  & 310 & 3.79& 	2 & 	14 &	95 &	135 &	64 \\ \hline
Bayes Theorem & 305 & 3.63 &	5 &	28 &	93 &	127 &	52 \\ \hline
RV Modeling  & 305 & 3.64	& 4	& 20	& 110	& 119	& 52\\ \hline
ML Fairness  & 300 & 3.65	& 6	& 25	& 100	& 107	& 62 \\ \hline
    \end{tabular}
    \caption{Student ratings of Real World problems on a scale of Strongly Disliked (1) to Strongly Liked (5).}
    \label{tab:ProblemAffinity}
\end{table}

General sentiment in long-form feedback (Q6-Q9) was consistent with the numerical data. Responses in long-form feedback was coded for sentiment: generally positive, negative, or neutral. Emotion had to be clear in the response to get a code; many comments did not receive any sentiment code. Neutral sentiment required a mention of weak/non-existent emotion  (e.g., ``it was fine'') or a combination of positive and negative sentiment in the same response. Q6 was asked twice (on the homework and in the post-survey), we only found 13 instances (DM: 5, PS: 8) where students had inconsistent emotions between the two times. The response was tagged with two different sentiment codes along with the inconsistent code. Far more positive comments (DM: 308, PS: 375) were registered than negative (DM: 41, PS:50) or neutral (DM: 17, PS: 34). Some students explicitly asked for more Real World Problems to be added to the homework (DM: 15, PS: 4).

Students regularly reported that problems were interesting or engaging (DM: 136, PS: 73). One student said ``There was potential motivation for me to do critical thinking, and the idea that there are problems required to be solve[d]/analyze[d] by theoretical knowledge made the learning process exciting.'' another ``It can activate my pas[s]ion to learn.''

\section{Impact on Learning} \label{sec:result4}
Student feedback on the problems showed significant potential for helping students learn. Many students reported that doing the problems helped them understand the course concepts better (DM: 108, PS: 134) with a few students saying that it didn't have an effect on their learning (DM: 18, PS: 31) or had a negative impact (DM: 2, PS: 0). One student said ``The Real World [P]roblems definitely helped conceptualize the more theoretical concepts, especially as they were more abstract'', another mentioned that ``Since we often needed to spend more time on finding related real-world examples and on thinking about how to correctly apply the theorem, I felt it helped leave stronger impression for the respective concepts, and this impression did deepen my understanding of the topics than those not associated with the questions'', and another said the Real World Problems ``motivate[d] me to keep brushing up on the concepts and exploring theoretical computer science even after the class has ended. This makes it easier to recall the concepts learned because they are connected to a relevant example.''

\section{Ethical Implications} \label{sec:result5}
Real World problems also gave students an opportunity to consider ethical problems of the use of course concepts in the real world. While no code looked for ethical observations separately, many students commented on the ethical implications we hoped they would discover.

In regard to the propositional logic problem, one student said:
``I can't believe this [court case] is a real event. Maybe every time a new law is made, we should ask mathematicians to analyze whether the law is rigorous enough in expression and scope of application.'' Another said that the problem ``brought to light the ethical dilemmas of using logic to talk about conflict involving people.''

In regard to the random variable modeling problem, one student said:
``It was fascinating to see how easily people either confuse or even intentionally misuse certain information yet still maintain its legitimacy because of mathematical calculations.''

In regard to the machine learning fairness problem, one student said: 
``This was one of the most thought-provoking problems, as it connected theoretical CS concepts with ethical considerations.''

At the end of PS one student said the Real World problems should be offered in future quarters because ``I think they provide an important perspective on how to look at the arguments having to do with probability in the real world. When I see an argument with probability now I am wondering, what went into this calculation? How much data did they collect? What does their wording actually mean? Are they misrepresenting this data in any way?''

% ========== Chapter 6
  \chapter{Limitations} \label{ch-6}
    
    \section{Cohort Effects} As DM is the first course restricted to CS majors, the student population can vary by quarter. For example, it tends to have more community college transfer students in Autumn than in Winter. Any such effects did not appear to significantly affect responses in the pre-survey (See \Cref{tab:SequenceChiSq} Rows 1,2), but it is still possible that the cohorts had similar initial responses but were differently open to changing their minds, for example. 
    \section{Social Desirability Bias}  
    Survey respondents can be biased toward giving answers that they think the pollster is hoping for \cite{grimm2010social}. Students may have been more likely to give agree responses, particularly in intervention quarters, where it was their instructor running the survey.
    To reduce potential impact, students were told that surveys were anonymous, and we would not look at results at all until after the term ended. 
    \section{Other real world applications in courses} Both courses (in both intervention and control offerings) cover real world and future CS applications of content in lecture, which could have impacted student opinions as well.
    \section{Survey fatigue} Given the number of spots we requested feedback, students may have been tired of surveys, particularly for those who took intervention quarters consecutively.

% ========== Chapter 7
\chapter{Best Practices} \label{ch-7}
\section{Distribute problems throughout the term} Many students reported appreciating a different kind of problem (DM: 19, PS: 14) as a break from the regular problems and found them easier (DM: 7, PS: 18). ``The real world problems are fun, offer a new perspective, and a change of pace. To exaggerate slightly, they are good practice without the feeling that it is `scary' or serious''. Others reported finding problems challenging (DM: 28, PS: 14) or mentioned confusion (DM: 50, PS:43)\footnote{Where comments related to particular parts or phrasing (36 of 93), we used comments to improve phrasing for future versions. Other comments on confusion were students identifying struggles with technical concepts (8 of 93) or mentioning issues inherent to open-ended problems (17 of 93). Sadly, a large number (26 of 93) of these comments were so short as to provide no more information than that the student felt confused.}. Spreading out problems during a term gives the benefit of a break to students enjoying them while minimizing the frustration for those who don't want them. Richard et al. made a similar recommendation for their inclusive combinatorics problems \cite{10.1145/3478431.3499330}).

\section{Keep problems as short as feasible} Real world problems will necessarily have more reading to provide context. Some students found the amount of reading off-putting (DM: 8, PS: 2) or simply reported that they spent a significant amount of time on the problem (DM: 11, PS:50). Despite this commentary, we did not find Real World Problems noticeably increasing reported time spent on homework; only one of seven Real World Problem (see \cref{tab:TimeReportedDM} and \cref{tab:TimeReportedPS}) was the plurality selection to the question ``Which problem did you spend the most time on?'' in homework-by-homework feedback. Problems shorter than a regular homework problem seem to be a good target. 

    \begin{table}
    \centering
    \begin{tabular}{|c|c|c|c|c|c|c|}
    \hline
    \textbf{Homework} & $n$ & \textbf{Median (hr)}  & \textbf{RWP Voted As Longest} & \textbf{Most Voted} & \textbf{RWP Votes} \\ \hline
    HW 1 & 306 & 4 & N/A & 136 & N/A \\ \hline
    HW 2 & 297 & 7 & Not Selected & 78 & 35 \\ \hline
    HW 3 & 300 & 5 & N/A & 121 & N/A \\ \hline
    HW 4 & 294 & 7 & Not Selected & 179 & 41 \\ \hline
    HW 5 & 285 & 9 & N/A & 223 & N/A \\ \hline
    HW 6 & 283 & 6 & Not Selected & 84 & 77 \\ \hline
    HW 7 & 288 & 7 & N/A & 88 & N/A \\ \hline
    HW 8 & 277 & 9 & N/A & 126 & N/A \\ \hline
    \end{tabular}
    \caption{DM HW Student Feedback; Students reported how long they worked on the HW and which problem took them the most time. We compare how may votes the Real World Problem got (if applicable) to number of votes for highest selected problem.}
    \label{tab:TimeReportedDM}
\end{table}

    \begin{table}
    \centering
    \begin{tabular}{|c|c|c|c|c|c|c|}
    \hline
    \textbf{Homework} & $n$ & \textbf{Median (hr)}  & \textbf{RWP Voted As Longest} & \textbf{Most Voted} & \textbf{RWP Votes} \\ \hline
    HW 1 & 334 & 7 & N/A & 108 & N/A \\ \hline
    HW 2 & 328 & 9 & Not Selected & 242 & 6 \\ \hline
    HW 3 & 300 & 10 & N/A & 133 & N/A \\ \hline
    HW 4 & 317 & 10 & Selected & 150 & 150 \\ \hline
    HW 5 & 318 & 6 & N/A & 132 & N/A \\ \hline
    HW 6 & 310 & 7 & N/A & 125 & N/A \\ \hline
    HW 7 & 309 & 10 & Not Selected & 193 & 7 \\ \hline
    HW 8 & 313 & 8 & Not Selected & 145 & 66 \\ \hline
    \end{tabular}
    \caption{PS HW Student Feedback; Students reported how long they worked on the HW and which problem took them the most time. We compare how may votes the Real World Problem got (if applicable) to number of votes for highest selected problem.}
    \label{tab:TimeReportedPS}
\end{table}

      \section{Provide options for open-ended questions} Including options that students can use for open-ended portions both makes expectations more clear for students finding their own and gives students an alternative option if they cannot find an example they like. Provided options were used by a majority of students, so they should be chosen carefully. Many students still chose to find their own examples and frequently reported struggling to find a real world example they could fit into problem constraints (DM: 21, PS: 46). While some students reported disliking the open-ended portions (DM: 2, PS:8), more reported liking them (DM:16, PS:47).

    \section{Clearly communicate expectations for solutions} Students expressed concerns about getting open-ended questions right (DM: 11, PS:15). Telling students explicitly that there are multiple correct answers, when grading is based on completion instead of correctness, or including suggested response length may help alleviate that concern. We experimented with providing sample full-credit solutions or full rubrics to students in advance; these were a significant effort without much explicit positive feedback from students, but might be helpful in some cases.

    \section{Keep Grading simple for TAs} Open-ended parts can pose a challenge for TA grading due to the variety of responses at scale. In courses with 300+ students, we found the following approach helpful 
    based on the Kuznetsova's and other TAs' experience grading. 
    We recommend that all open-ended portions of a submission are graded by the same TA (rather than having TAs grade one sub-part of all submissions). This scheme allows TAs to have the full context of a response and avoids repeating work when grading submissions by parts. We also found that having the open-ended and reflection potions of the Real World Problems graded on completion (as opposed to correctness) is best. This simplifies grading to only a few rubric items for clear deductions (e.g., missing a required part).
    
    \section{Consider potential grade implications} The homework averages in our courses tend to run very high (medians approximately $90\%$ on many assignments). Anecdotally, a few students avoided finding their own example (of greater interest to them), and used our examples instead in an effort to avoid losing any points. A common comment in positive feedback about the problems was that students found them easier or less stressful than regular problems (DM: 7, PS: 18), which may have helped their popularity among some students.

% ========== Chapter 8
\chapter{Conclusion} \label{ch-8}
Many concepts are amenable to the structure we aimed for with our Real World Problems: a small technical problem, an open-ended search for an application of the topic, and a reflection on the connection. The structure may be useful, even for courses where students don't frequently ask why they are learning the content.

 In theory courses (and abstract courses more generally), ethics content can be uniquely impactful. Problems will inherently connect the course to the real world (which will make the course feel applicable), while also integrating the ethics lessons the instructor desires. A larger library of these problems would be worthwhile, as they could address both goals at once.

Future work should examine the fraction of students who are unmotivated for theory courses in more contexts, as there seems to be some tension in our data on the topic between quantitative results in Q1-Q4 showing relatively small changes of increase and some decrease over the term for both intervention and control terms, but the qualitative feedback having frequent mentions of new observations of relevance. We hypothesize that social desirability bias could play a role in high relevancy (especially in the pre-surveys, where opinions may be less fixed), but resolving this tension is beyond the scope of this thesis. It may be worth running similar surveys with a more carefully designed instrument and in more contexts as part of a detailed education research paper. 

Real World Problems are a simple-to-implement method of strengthening student belief in the relevance of the content. Students enjoy them, and they allow students to engage with  content in new ways.
%
% ==========   Bibliography
%
% \nocite{*}   % include everything in the uwthesis.bib file
\bibliographystyle{plain}
\bibliography{uwthesis}
%
% ==========   Appendices
%
\appendix
\raggedbottom\sloppy
 
% ========== Appendix A
 
\chapter{Real World Problem - Discrete Math\\ Homework 2 - Legal-ese} \label{DM-HW2}

Computer scientists aren't the only people who need to precisely interpret long English sentences. Lawyers and judges need to interpret long (but, hopefully, precise) sentences in law codes. These groups sometimes run into the same types of problems we have experienced in interpreting these sentences. In fact, a U.S. Supreme Court case in 2024 involved a disagreement on how to interpret a statement that could have been (maybe \emph{should} have been) put in propositional logic.  

The case was about a provision in a particular law, giving lighter sentences to defendants with less criminal history. It said (in part)

\noindent (*) \texttt{A defendant is eligible for relief [a lighter sentence], if the defendant does not have---
        \begin{itemize}
            \item[(a)] Condition A;
            \item[(b)] Condition B; and
            \item[(c)] Condition C.
        \end{itemize}
}
Where conditions A, B, and C were specific measures of prior criminal convictions.
A specific defendant (Pulsifer) did not have Condition C, but did have Conditions A and B. So, could he get a lighter sentence? 

Pulsifer said he could! If we let $e$ be the proposition ``the defendant is eligible for relief'', and $a,b,c$ be propositions corresponding to the conditions in the law, then the law said $[\neg(a\land b \land c)] \rightarrow e$. Since $c$ is false for him, the hypothesis is true, meaning that he must be eligible for relief. 

But prosecutors said he was not necessarily eligible. They thought the law said $[\neg a \land \neg b \land \neg c] \rightarrow e$. 

\begin{enumerate}
    \item Take a moment to look at the English sentence at issue (the sentence in \texttt{monospace} after the (*) above) and look the two interpretations in propositional logic. Notice what the interpretations have in common: both think we have an implication, both combine $a,b,c$ with ANDs, but they disagree on how ``not'' relates to the conditions: does it apply to the combination of all three, or does it relate individually to each one? Try to put yourself in each party's mindset and see how the English sentence could plausibly mean what they claim. \\
    If you don't feel like you understand the two interpretations, you might want to read \href{https://www.oyez.org/cases/2023/22-340}{this case summary} or Section II of the \href{https://www.supremecourt.gov/opinions/23pdf/601us1r07_19m1.pdf#page=10}{Supreme Court's opinion}, which have more details on the law. \\ 
    Don't write anything for this part. [0 points]

    \item At the Supreme Court argument, Pulsifer's lawyer said ``The government [prosecutors] needs `AND' to mean `OR''', but in propositional logic, the prosecutors still have ``AND''s.\\
    Convert the prosecutor's propositional logic to replace the $\land$s with $\lor$s and show your work with a chain of equivalencies (citing the appropriate rules). For this part, you are allowed to apply associativity, commutativity, and double negation without explicitly writing the step. [4 points]\\
    \textbf{Hint:} If you've done it right, the structure should look a lot like Pulsifer's interpretation of the law, just with OR's instead of AND's. That is to say, if you agree with where Pulsifer's lawyer put the parentheses, the alternate interpretation does want ``AND'' to mean ``OR.''
    \item Try to write a better English sentence than Congress did---write one that can correspond to Pulsifer's interpretation \textbf{only}, not the prosecutors' interpretation. You probably will want to use additional words. Phrases like ``simultaneously'' or ``everything in this list'' or ``at least one thing in this list'' can help make a sentence like this more clear; you might also try moving the ``not'' to be closer to each of the conditions. Alternatively, taking a contrapositive could help! [3 points]
    
    \item Do the same as the previous part but for the prosecutors' interpretation. Again, adding words or taking the contrapositive may be helpful to express the logic more clearly in English. [3 points]

    \item Take a moment to appreciate having propositional logic! It's hard to write an English sentence conveying either interpretation that can't be confused with the other one. But notation lets us differentiate between the two (comparatively) easily. You do not have to write anything for this part [0 points]

    \item What do you think is the ``right'' interpretation of the law? Pulsifer's? The prosecutors'? Is the law ambiguous with no clear interpretation? What should a court do in a situation like this?
    We won't grade your answer on whether we agree with you, just whether you've addressed the questions (3-4 sentences, but feel free to write more).\footnote{Many expert lawyers and judges have come down on both sides of this argument, there's plenty of room for disagreement!} [2 points]
\end{enumerate}

\chapter{Real World Problem - Discrete Math\\ Homework 4 - Logic Beyond CS} \label{DM-HW4}
Find an example of someone making an argument in real-life, and determine if they made an error involving mixing up an implication and its converse. This could be in the news (like a politician or opinion piece), in culture (like a TV clip or an advertisement), or just something your friend said. If you can't think of one, you may use one of the examples in the section below.
\begin{enumerate}
    \item Write down a quote of the person making the argument. 
    \item If you can link us to a source, do so here (don't worry about formatting). If you can't (because there's no record of the statement), just say that.
    \item From the quote you have in (a), respond to the following: \\ (i) Define propositions that they are dealing with and translate what they say to propositional logic.\\ (ii) What givens are they asserting and what is their conclusion (i.e. what do they intend to argue for)? \\
    (iii) Are they making an error? If so explain why. If not, what inference rule(s) could you apply to formalize their argument?\\
    Since people tend to speak informally (rather than in precise logical arguments), you may have to read between the lines a bit when doing this part.
    \item Suggest replacing one (or more) of the implications with their converse in the propositional logic you have in (c). With the change, can one reach their desired conclusion without making any logical error? (1-3 sentences) 
    \item Do you think the converse(s) you inserted in the previous part are true? Or at least ``often true''?\footnote{For example, it might not be true that ``If I have my umbrella, then it is raining'' (since I also bring my umbrella when it snows), but snow is so rare that the implication would `often' be true.} Explain in 1-3 sentences.
\end{enumerate}

% \subsubsection*{Some options}
\noindent \textbf{Some options}

You're encouraged to keep your eyes out for this error in real life! Or to think about places you might have seen it. If you cannot find one, you might choose one of these options instead:
\begin{itemize}
    \item This cartoon about logical penguins:
    \begin{center}
    \includegraphics[scale=0.5]{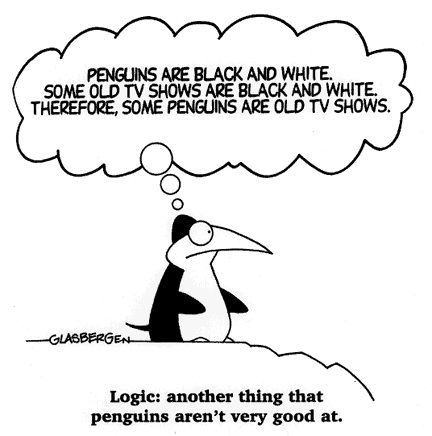}.\\
    \end{center}
    \item \href{https://www.youtube.com/watch?v=fm2W0sq9ddU}{This clip} from The Simpsons.
\item \href{https://youtu.be/zwsmqZLCKPE?t=53}{This clip} from Sesame Street.
    \item The clip from the childrens' show ``If you give a mouse a cookie'' available on \href{https://github.com/annak4397/UW-CSE-Real-World-Problem-Research-Project/blob/main/problems/updated-problems/resources/If%20you%20give%20a%20mouse%20a%20cookie%20(Clip).mp4}{Github}.
\end{itemize}

\noindent \textbf{At UW [6 points]}

\textbf{Note for instructors:} this is a stand-alone part that can be omitted to create a shorter problem.

Take (and attach) a photo of some sign (preferably at UW, but we will accept any signs you see in your day-to-day life), break the contents of the sign into atomic propositions and provide the logical translation of this sign. Please also mention where you found this sign (so perhaps if any of your TA's are bored, we can do a logical sign trek :D).\\
Example: 
\begin{center}
    \includegraphics[scale=0.25]{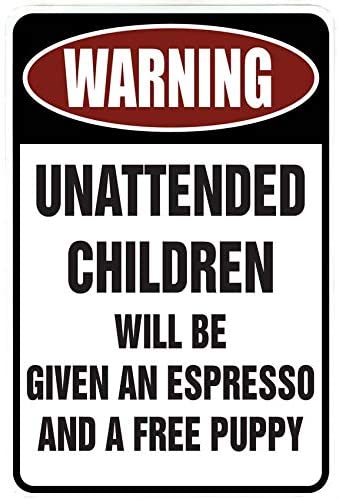}.\\
\end{center}

We can represent this with the atomic propositions :\\
$p: $ One is an unattended child.\\
$q: $ One will be given an espresso.\\
$r: $ One will be given a free puppy.\\
$p \rightarrow q \land r$
\begin{enumeratequestions}
    \item Attach a photo of some sign / iconography you have found.
    \item If you feel comfortable, mention where you found this sign.
    \item Define atomic propositions.
    \item Translate the contents of the sign into logic using the atomic propositions defined above.
\end{enumeratequestions}

\chapter{Real World Problem - Discrete Math\\ Homework 6 - Some Simple Code} \label{DM-HW6}
Computer scientists write code, but if you want other people to use your code, you'll have to be able to explain why it works. In future classes (especially CSE 421), you'll do that with a proof.\footnote{In the real world, you won't do a full proof (unless you become a researcher), but you will still have to clearly explain what's going on in your code, and a proof is a good way to practice a careful explanation}

Let's take an example of code that returns the product of the first $n$ even integers for inputs of $n > 0$.\footnote{We took this coding problem from \texttt{Practice-It}; It was authored by Whitaker Brand (on 2019/09/19)}

\begin{lstlisting}
public static int multiplyEvens(int n) {
    if (n <= 0) {
        throw new IllegalArgumentException();
    }
    if (n == 1) {
        return 2;
    }
    return (2 * n) * multiplyEvens(n-1);
}
\end{lstlisting}

\begin{center}
\begin{tabular}{ |c|c|c| } 
 \hline
 Call & Output & Reason \\ 
    \hline
\texttt{multiplyEvens(1);} & 2 & 2 = 2 \\ 
 \texttt{multiplyEvens(2);} & 8 & 2 * 4 = 8 \\ 
 \texttt{multiplyEvens(3);} & 48 & 2 * 4 * 6 = 48 \\ 
 \texttt{multiplyEvens(4);} & 384 & 2 * 4 * 6 * 8 = 384 \\ 
 \hline
\end{tabular}
\end{center}

In this problem, you will write a correctness proof that will show that this code always returns the product of the first $n$ even integers for inputs of $n > 0$. In other words that $$\texttt{multiplyEvens(n)} = \prod_{i = 1}^n 2i$$
Capital pi ($\prod$) is the product symbol which represents repeated multiplication, similar to $\sum$ which is for repeated summation. For example, $\prod_{k=1}^3 4k = 4\cdot 8 \cdot 12$. 

\begin{enumerate}

\item Prove that the code produces the desired output.
You must use induction for this problem. Be sure to start by defining your predicate $P()$. [20 points]

\item Take a moment to reflect on the structure of the proof and the code. The code has a base case and a recursive case, which relies on the result for input $n-1$ to calculate the answer for input $n$. The proof will have a base case, and an inductive case from $k$ to $k+1$. You do not have to write anything for this part [0 points]

\item Imagine you were tasked with convincing someone the code snippet works. What would you do (write this proof? Run some test cases? A combination of those? Something else?)? How would your strategy change if this code snippet had a different structure (e.g., no recursion)? (3-4 sentences, but more is okay) [3 points] 
\end{enumerate}

\chapter{Real World Problem -  Probability and Statistics\\ Homework 2 - Is this a pigeon?} \label{PS-HW2} \footnote{This problem is an adaptation of a problem written by Shreya Jayaraman in \href{https://courses.cs.washington.edu/courses/cse312/23su/}{Summer 2023}.}
The pigeonhole principle is surprisingly powerful and can be used to prove a variety of things - find one application of the principle that interests you (it can be CS-related or not, serious or not; the only requirements are that (1) you must find it interesting and (2) we shouldn't have proved it already in this course) and explain the proof to us. Your solution needs to include these three parts:
\begin{enumerate}
    \item Your source and some background for the application.
    \item Relate your proof to the theorem definition from lecture - what are the pigeons and what are the pigeonholes?
    \item What does the theorem tell us about them?
\end{enumerate}

\noindent \textbf{Some suggestions}

We hope you'll think of something on your own! If you can't, here are some ideas that you can use:
\begin{itemize} 
\item Husky Den Food Court options
\item UW Course Conflicts (e.g., based on times of day; you can't do number of courses you take per quarter since we did that one in class).
\item Olympic Sports (e.g., finish times of races)

\end{itemize}

\chapter{Real World Problem -  Probability and Statistics\\ Homework 4 - Bayes Theorem} \label{PS-HW4}
The tools of this class are useful to computer scientists, but many of them are useful beyond just ``classic'' computer science. In this assignment you'll consider an application of Bayes' Rule in the real-world.

We will consider the use of DNA evidence in criminal trials. A full discussion of DNA evidence would require a discussion of many issues\footnote{Among others: under what circumstances DNA samples be taken from people and/or stored in databases.} -- for this assignment, we are going to limit ourselves to just how information about DNA tests should be communicated to juries. 

This assignment is a mix of technical tasks (appropriately applying theorems) and non-technical ones (considering tradeoffs between various real-world effects and groups). The technical aspects can be ``right'' or ``wrong'', but the non-technical aspects are unlikely to be simply ``right'' or ``wrong'' -- we won't have to \textbf{agree} with the non-technical aspects of your analysis to consider them a good analysis. Our evaluation will be based on how well they connect to the technical aspects, as well as the depth of reasoning demonstrated.\footnote{For example, trying to calculate a probability and getting $1.2$ for an answer would involve a technical mistake. Saying ``Witnesses shouldn't say the DNA evidence is reliable, because I saw an episode of CSI where it wasn't reliable.'' is not good reasoning for {this assignment} because it does not connect to the technical aspects of the problem. Saying ``DNA evidence should be allowed as long as the Bayes factor is at least 100'' relates to technical aspects and is considered good analysis whether or not we agree with you on ``Bayes factor at least 100'' being the right place to draw the line between allowable or not.}

\noindent \textbf{Collaboration Policy:}

For the work in \cref{subsec:yourArg}, you are to conduct your own search and analysis for this assignment. While you may get feedback from other students on your writing, you cannot just use the results of another student's search. 

% \subsection{Bayes in Court}
\noindent \textbf{Bayes in Court}

DNA evidence has been used in court cases for decades. Over time some common patterns of (dubious) argumentation have emerged, which you'll analyze in this problem.

 Consider the following scenario:
 
 A crime is committed somewhere in Seattle. No witnesses were at the crime, but there was blood left at the scene which had DNA extracted from it.
 The DNA was run against the 13 million DNA samples on file with the FBI. There was one match: a person who lived in Tacoma at the time of the crime.
 
 You know the following facts about the DNA test that was run:
 \begin{itemize}
     \item The false positive rate of the test is $\frac{1}{10,000,000}$.
     \item The false negative rate of the test is $\frac{1}{100,000,000}$.
 \end{itemize}
    The prosecutor argues as follows
    \begin{quote}
        The DNA match with the blood on the scene is strong. There is only a $\frac{1}{10,000,000}$ chance that the defendant is innocent (after all, the test only has a $\frac{1}{10,000,000}$ rate of failure) -- certainly not a reasonable amount of doubt. You must vote to convict. 
    \end{quote}
Let $T$ be the event of a positive test, and $G$ be the event that the defendant is guilty.

\begin{enumerate}
        \item[(a)] In terms of $G$ and $T$, what probability or conditional probability is the prosecutor describing with their phrase ``the chance the defendant is innocent, knowing about the test''? [1 point]
    
        \item[(b)] What probability or conditional probability does the $\frac{1}{10,000,000}$ come from? [1 point]
        
        \item[(c)] Describe the prosecutor's error concisely (2-3 sentences). [2 points]
        
    \end{enumerate}

      The defense attorney argues as follows: 
     
     \begin{quote}
        The test isn't as good as it sounds. If we ran the test on all 330,000,000 people in the country, we'd have 33 innocent people come up with positive tests. The true probability of my client being guilty is only about $1/34$.
     \end{quote}  

     Recreate the Bayes' Rule application that the defense attorney is using
     \begin{enumerate}
         \item[(d)] What prior is being used and what is the assumption being made by the defense? Recall the ``prior'' is the probability of the event you're hoping to analyze \textit{prior to} running the test. Your answer here should include both a number and where it came from. [2 points]\\
         \textbf{Hint:} What is the sample space that the defense is referring to?

         \item[(e)] Now use Bayes' rule to confirm that (starting from that prior), the calculation is correct. [2 points]

     \end{enumerate}

     Now choose a new prior. What is \textbf{your} estimate of the probability the defendant is guilty? You can use either (or both) of the bullets below. If you use neither bullet, you must incorporate some other information and have something different from what the prosecutor and defense attorneys said. Since this is \textbf{your} estimate, there are many possible answers! We aren't grading whether we get the same answer, we're grading whether you have a correct application of reasonable assumptions.
     \begin{itemize}
         \item The 13 million DNA samples in the database are not from a random section of the population, but they do come from people across the whole U.S.
         \item The Seattle metro area has about 4 million people. 
     \end{itemize}
     
 \begin{enumerate}
     \item[(f)] What is your prior (i.e., the probability the defendant was guilty before you ran the test)? Briefly explain where it comes from. [1 point]
     \item[(g)] Do a Bayes Rule calculation to give your estimate of the guilt of the defendant.[2 points]

     \item[(h)] Name at least one limitation of your estimate (something you haven't accounted for that you would have liked to, or more information you would have liked about the scenario)? (2-3 sentences) [2 points]

 \end{enumerate}

\noindent \textbf{Make Another Argument} \label{subsec:yourArg}

\textbf{Note for instructors:} this is a stand alone part that can be added or omitted based on how long you want the Real World Problem to be.

In this part, you'll use an application of Bayes Rule to make an argument about whatever real-world scenario you would like.

Your scenario can be close-to-home (say something about an RSO you're involved in), a political issue, or anything else, as long as it's based in the ``real-world''\footnote{We will be quite lenient about what counts as real world -- the hope is that you will pick something you care about. If it's just the probability that the second and third card of a deck of cards are the same value, it's probably not ``real-world.'' But if you're an avid poker player, and you want to use Bayes' Rule to analyze a particular game scenario, that would definitely count. }.

You are allowed (and encouraged!) to do your own research toward this question, but can also fall back on reasonable estimates. 

\begin{enumerate}
    \item Define events $A$ and $B$ on which you'll apply Bayes' Rule (along with any other events you need). [2 points]
    \item State probabilities (or probability estimates) for three of the four quantities you need to use Bayes' rule and apply Bayes' rule. [1 point]
    \item For those estimates, either cite a source for the numbers that you think is reliable or give a justification for your estimate. [1 point]
    \item  Apply Bayes' rule using the probabilities from part (b) [4 points]
    \item What is your takeaway from this calculation? This needs to be more than just restating what your calculation in part b found. [2 points]
    \item Discuss at least one limitation of your calculation/application (e.g. factors that didn't go into your estimates, or assumptions you are making that might not be correct). [2 points]
\end{enumerate}

\noindent \textbf{Some Ideas}
We hope you’ll think of something on your own! If you can’t, here are some you might think about:
\begin{itemize}
	\item We saw in class that routine medical tests can lead to false positives/negatives. Some tests you might consider looking into are over-the-counter pregnancy tests, colon cancer tests, and paternity tests.
	\item How hard should \href{https://en.wikipedia.org/wiki/CAPTCHA}{Captchas} and other ``I'm not a robot'' tests be to stop the robots from random guessing, but allow through fallible humans?
    \item How reliable is the rain prediction in weather apps for Seattle?
\end{itemize}

\chapter{Real World Problem -  Probability and Statistics\\ Homework 7 - Modeling Assumptions} \label{PS-HW7}
The tools of this class are useful to computer scientists, but many of them are useful beyond just ``classic'' computer science. In order to use the powerful tools of probability, we need to make assumptions to let our mathematical tools model the real world. Things like ``this coin is perfectly fair'' or ``the coin flips are all independent'' are usually not perfectly true.\footnote{e.g., if you flip a coin repeatedly, the result of the last flip is probably how the coin will appear on your hand before you flip it, which will make the results not quite independent.} Indeed, occasionally these assumptions are ways that people ``lie with statistics'' or provide evidence for claims that aren't actually true.

In this question, you will critique the modelling assumptions made in an analysis.

Find an analysis (e.g., via a blog post/article) that uses probability and statistics tools you're familiar with from this course. By ``analysis,'' we mean any estimate of a ``real-world'' probability, along with the assumptions that lead to that number. You might want to look at the examples in section \ref{sect:ideas} for what we mean.

We expect most of the answers to this section will be short (2-3 sentences), but you are free to write more if your resource is more complicated.
\begin{enumeratequestions}
    \item Provide a link to (or somehow let us access) the analysis you're critiquing. [3 points]
    \item What is the fundamental claim of the analysis? I.e., what conclusion do they draw at the end of their analysis? [3 points]
    \item What modelling assumptions do they use? (For example, do they assume some occurrences are independent? Do they assume a set of events all have equal probability? Do they assume they know the probability? Do they use a variable from the zoo?) [3 points]
    \item Do you believe their analysis is correct? If so, choose an assumption from C and explain why you agree with it. If not, what assumption would you change?\footnote{\textbf{Note for instructors:} prior versions asked students to change an assumption, redo the calculation, and compare the results to see if the conclusion would change. Adding that back into the problem would expand it, but would make it more difficult (because the example computation needs to be flawed, which can be harder to find).} [3 points]
\end{enumeratequestions}

\noindent \textbf{Some Ideas}
\label{sect:ideas}

You are free (and encouraged!) to find your own examples outside this list if you have a topic you are passionate about, but if you can't think of anything, you may use any of these as starting points. 
In many cases, there are already critiques of poor statistical/probability analyses online -- it's ok to look at these critiques, as long as you tell us if you're using any and still do the new probability calculation independently and put everything in your own words.

\begin{itemize}
    \item \href{https://www.reddit.com/r/TheSilphRoad/comments/9738aq/question_about_probabilities_shiny_rates/}{Some} think that the probability of encountering at least one shiny Pokémon does not change with seeing more Pokémon. 
    \item Richard Lustig, 7-Time Lottery Winner, Gives Tips On Winning The Powerball Jackpot. You might find that his advice doesn't make the best assumptions. See article \href{https://www.huffpost.com/entry/richard-lustig-7-time-lottery-winner-how-to-win-powerball-jackpot_n_2203717}{here}.
    \item Is \href{https://www.weather.gov/safety/lightning-odds}{this} an accurate estimate of the probability of being struck by lightning in the US? Could being in different states have an impact on this estimate?
    \item Every year millions of people predict the outcomes of the NCAA men's basketball tournament. It is commonly said that the probability of a perfect bracket is $\frac{1}{2^{63}}$, (since there are $2^{63}$ ways the $63$ games could play out) and therefore no one will ever predict a perfect bracket. (Yes, Robbie has a blog post about this one) \href{https://www.usatoday.com/story/gameon/2013/03/19/ncaa-tournament-perfect-bracket-odds-quintillion/1999795/}{Here is a source using that number}
\end{itemize}

\chapter{Real World Problem -  Probability and Statistics\\ Homework 8 - ML Fairness} \label{PS-HW8}
    
One of the most popular (and influential) applications of the tools of this course to the real world has been in machine learning. 
You have probably (hopefully!) heard that ML applications have had unintended effects in the real world, especially around fairness. In this assignment, we will use the tools of probability to get a taste of what fairness analysis looks like. Our specific goals are:
\begin{itemize}
    \item Practice some old concepts (right before the final) by applying them in a new context.
    \item Practice converting between notation and a more intuitive understanding.
    \item See that probability and conditioning are useful formal representations of what it might mean to be fair.
    \item Realize that ``fair'' has more than one reasonable definition, and you can't necessarily have all of them!
    \item Get a sense that fairness is complicated!
\end{itemize}

This will \textbf{not} be a full introduction to fairness in ML; the topic could easily fill an \href{https://courses.cs.washington.edu/courses/cse599m/20wi/}{entire course}. This assignment is intended to help you realize that the topic is important, requires careful attention to understand, and that it is a possible application of the tools you learned in this course.

When thinking about ML systems, we need to be careful about what it means to be ``fair.'' A definition that leads to equitable outcomes in one setting may have the opposite effect (setting in stone historic inequality) in another scenario. The actual definition of ``fair'' to pick is thus critical -- we can't know if we're being fair or not if we don't know what fair means! And (as we see next), definitions of fair that might be reasonable in one context or another are sometimes even contradictory! You will see three definitions of ``fairness'' that have been used in the literature. We'll then consider one potential area of application for these definitions -- you'll choose which of the definitions you believe is appropriate for that setting. 

There are other questions one must ask about an ML system before using it (should one use an ML system at all? If so what data is appropriate? If so should a human be in-the-loop and how?) As we said right at the start of this assignment these issues are subtle and require careful thought! We can't handle every aspect in a single assignment, so for this one we limit ourselves to just defining fairness.

The basic operation of machine learning is to take in a large set of data, find patterns in that data (using statistics), and make predictions about future data based on the given data. 

\noindent \textbf{Three Definitions of Fairness [6 points]}
    
    ML Fairness is usually discussed in the context of disparate outcomes across social groups, and often in terms of effects on \href{https://en.wikipedia.org/wiki/Protected_group}{protected classes} (e.g., based on a person's sex or gender or race---the linked Wikpedia article has a longer list of examples). 
    
    To make the examples more concrete, imagine that our ML application is for a bank deciding on making loans. Our ML system will look at historical data of who repaid loans, and use that to predict whether a new person will repay their loan.

    Suppose we choose a loan applicant at random:
    Let $L$ be the indicator that our ML system says our applicant should get a loan, let $A$ be the the indicator that the applicant has the protected-class attribute which we are examining, and let $Y$ be the indicator that the person would truly successfully repay a loan.\footnote{To make it possible to know $Y$ (without using the system to decide on real loans before knowing whether it works) ML systems are usually evaluated on a ``test set.'' A dataset gathered in advance where the true answers (the $Y$'s) are known, but the data was not used to design the model.}

    A ``fairness criterion'' is a possible definition of what constitutes ``being fair.'' In this assignment we'll see three definitions of ``fair'' that have been used in the fairness literature. For example, ``precision parity'' is defined as $\mathbb{P}\left(Y=1 | L=1,A=1 \right) = \mathbb{P}\left(Y=1 | L=1,A=0\right)$.
    That is, if the ML system says our applicant should get a loan, the probability that this applicant will repay the loan remains the same regardless of whether they have our protected attribute or not.
    Intuitively, the goal of this definition is to ensure that the ML system does not give more ``benefit of the doubt'' to one group than another (by ensuring that both groups have the same fraction of repaid loans). 

    \begin{enumeratequestions}
        \item Another possible fairness criterion is ``true positive parity'': $\mathbb{P}\left( L=1 | Y=1, A=1\right) = \mathbb{P}\left( L=1 | Y=1, A=0\right)$. Give a (1-2 sentence) intuitive description of what this criterion is trying to achieve. [3 points]
        \item A third possible fairness criterion is ``false positive parity.'' This definition seeks to ensure that among those who do not repay their loan, the probability of our system wanting to give them the loan is equal whether they have the protected attribute or not. Intuitively, the goal is not to give ``unfair'' benefits (loans unlikely to be repaid) to people without (or with) the protected attribute.

        Give the mathematical notation for false positive parity. [3 points]
    \end{enumeratequestions}    
    \noindent \textbf{You Can't Have it All [6 points]}
        
        A theorem\footnote{proven in the last 10 years!} says that an ML system cannot achieve all of precision parity, true positive parity, and false positive parity at the same time unless some unrealistic conditions are met\footnote{Specifically, those exceptions are: if the ML system never makes a mistake (which never happens at scale---there are simply too many variables for real people to understand them all perfectly) or if $\mathbb{P}(Y=1 | A=0) = \mathbb{P}(Y=1|A=1)$, that is the actual repayment rate of loans is equal regardless of the protected attribute (which, since protected classes are defined for protection because of historic disparate treatment, is uncommon in real data).}.

        \begin{enumeratequestions}
            \item For concreteness, pick an attribute which you want to ensure is treated fairly for this section. You can pick anything from the list of \href{https://en.wikipedia.org/wiki/Protected_group}{protected classes on wikipedia} or another attribute you pick. Tell us which one you picked.
            \item For the loan context, suppose you could make the system meet exactly one of the three fairness criteria above. Which of the three definitions of fairness would you want to ensure your ML system meets, and why? Any of the three can be chosen, but we want your answer to show an understanding of the definition and relate to the real-world; we won't grade based on whether we agree with you, just whether you've discussed your reasoning. [3-4 sentences]
            \item Choose one of the other fairness criteria (not the one you chose in the last part). Think of a scenario (something other than ``deciding on whether to award loans'') and an attribute where you think that idea of fairness would be the most appropriate of the three. Explain why in that scenario your choice is a good one. [3-4 sentences]
        \end{enumeratequestions}

\noindent \textbf{In the news [8 points]}

\textbf{Note for instructors:} this section is a stand-alone problem, which can be omitted for a shorter problem.

Find an news article about a particular instance of bias in a machine learning (or artificial intelligence) system. 
We expect the questions in this section can be answered in 1-2 sentences each.
\begin{enumeratequestions}
    \item Tell us the title of the article, the source (e.g., ``New York Times'') and give us a link to it (don't worry about pretty formatting).
    \item What is the ML application trying to learn (e.g. ``housing prices'' or ``recidivism rates'')? What group does the article suggest is being treated unfairly? 
    \item An ML application can be (roughly) grouped into 3 steps:
    \begin{enumerate}
        \item Data acquisition -- collecting the data that will be the basis of the system
        \item Model selection -- choosing which ``features'' of the data to use in the system and how those data points will be combined (e.g. for a house pricing model, one might decide whether or not to include the number of rooms, square-footage, and time that it has been on the market without selling as `features' to predict the selling price)
        \item Use of prediction -- once the model is trained, it will be used to make predictions (the model thinks this house that just went on the market will be worth $Y$ dollars), and those predictions will then be used in the real-world (e.g. a company might decide whether to invest in a property as a result of comparing the model's estimate of its value to its current price)
    \end{enumerate}
    Does the article you read try to place the blame on any of these steps? If so, which one(s)?
    \item Does the article state what the author thinks ``fairness'' would be? If not, is it implied? The definition might match the definitions in this assignment or be something different (there are other fairness criteria we haven't covered). 
    \item For this application, what definition of fairness do you think is appropriate (this could be one of the three definitions above, or it could be another definition). Your definition here can be phrased intuitively or in the language of probability (you don't have to list both).
\end{enumeratequestions}

\chapter{Codebook} \label{Codebook}
\begin{longtable}{|p{2in}|c|c|p{2.5in}|}
    \hline
       \textbf{Code}  & $n$ \textbf{in DM} & $n$ \textbf{in PS} & \textbf{Example Quote}\\ \hline    \endfirsthead
       \hline \textbf{Code}  & $n$ \textbf{in DM} & $n$ \textbf{in PS} & \textbf{Example Quote}\\ \hline \endhead
       Anxious or nervous at the start of the course. & 3 & 1 &  ``anxious about starting classes at UW in general''\\ \hline
       Application of concepts is more clear with the Real World Problem(s). & 25 & 0 & ``made the potential applications of concepts learned in this class clearer'' \\ \hline
       Appreciated having a different kind of problem. & 19 & 14 & ``They changed up the homeworks and made them more varied'' \\ \hline
       Asked for more Real World Problems to be added to homework. & 15 & 4 & ``Please have more of these kinds of questions in future homeworks.'' \\ \hline
       Believe that other students see the course content as irrelevant. & 5 & 1 & ``Real World problems help connect how proofs/discrete math can be tied back into the real world. There seems to be a general sentiment to some that 311 is irrelevant to CS  so having these problems may help make it more clear as to how discrete math in CS is interconnected.'' \\ \hline
       Curious to learn more at the beginning of the course. & 7 & 2 & ``I am curious to see the applications of the class material to practical applications.'' \\ \hline
       Deeper reflection on concepts or the real world. & 30 & 27 & ``The world does not operate in binary. There is lots of gray area.'' \\ \hline
       Did not think that the Real World Problem(s) related to the real world.& 1 & 0 & ``I dont feel like it relates to the real world. This problem [CSE 311 HW6 Some Simple Code] is hard to relate.'' \\ \hline
       Didn't want to come up with their own example for Real World Problem(s). & 2 & 2 & ``I disliked that we had to find our own example'' \\ \hline
       Disliked the open-ended portion of Real World Problem(s). & 2 & 8 & ``I think they would be better if they had more constraints or a stronger connection to course content instead of being open ended web searches.'' \\ \hline
       Enjoy(ing/ed) class. & 18 & 0 & ``I really enjoyed this class!'' \\ \hline
       Felt that the Real World Problem(s) lacked relevance to the course content. & 1 & 6 & ``I wasn't sure how applicable/relevant it was to the material we learned in class.'' \\ \hline
       Felt that the Real World Problem(s) were just extra work. & 0 & 1 & ``I understand that questions like these are important but they seem like extra work compared to how much else we need to do in class'' \\ \hline
       Felt that the Real World Problem(s) were unnecessary. & 2 & 0 & ``They're unnecessary in my opinion.'' \\ \hline
       Felt the content will play a role in their future career. & 1 & 1 & ``Important in real life and car[e]er to hone your intuition'' \\ \hline
       Felt the Real World Problem(s) were too contrived. & 3 & 0 & ``I thought most of the real world problems were sort of forced.'' \\ \hline
       Found Real World Problem(s) challenging. & 28 & 14 & ``It was hard.'' \\ \hline
       Found Real World Problem(s) easier. & 7 & 18 & ``They feel easier and lower-stakes than the rest of the problems but are still super helpful to my learning'' \\ \hline
       Found the amount of reading with Real World Problem(s) off-putting. & 8 & 2 & ``There is too much reading.'' \\ \hline
       Had suggestions for improving Real World Problem(s). & 10 & 22 & ``I think those kind of problem could be a very good candidate for extra credit problem (if we can have bonus points for writing/reflecting more in depth).'' \\ \hline
       Inconsistent comments about Real World Problem(s). & 5 & 8 & ``After problem: I enjoyed the application to the real world for this problem. End of course reflection: Was too vague.'' \\ \hline
       Liked having examples to use for Real World Problem(s).& 4 & 8 & ``It was helpful to have examples of similar arguments listed.'' \\ \hline
       Liked the open-ended portion of Real World Problem(s).& 16 & 47 & ``I liked searching up real world examples, I thought it was a good way to get more engaged with the material!'' \\ \hline
       Mentioned confusion with regards to Real World Problem(s). & 50 & 43 & ``trouble figuring out what exactly I was supposed to do for a few of the questions. The wording just made me confused'' \\ \hline
       Negative comment about Real World Problem(s). & 41 & 50 & ``It's a good problem I just don't like it'' \\ \hline
       Neutral comment about Real World Problem(s). & 17 & 34 & ``I thought it was fine.'' \\ \hline
       Positive comment about Real World Problem(s). & 308 & 375 & ``I liked them overall.'' \\ \hline
       Real World Problem(s) caused students to have a small change in perspective about the real world. & 52 & 5 & ``Let me think of things in a different way that I normally would.'' \\ \hline
       Real World Problem(s) caused students to think about the course concepts differently. & 23 & 18 & ``They made me think about the concepts differently, applying them to other areas rather than just brute-force knowing the basic skills.'' \\ \hline
       Real World Problem(s) did not have an effect on their learning. & 18 & 31 & ``It just made it more enjoyable  did not affect understanding.'' \\ \hline
       Real World Problem(s) had a negative impact on their learning. & 2 & 0 & ``I think they were interesting  but was more confusing than helpful.'' \\ \hline
       Real World Problem(s) had no effect on their view of the real world. & 37 & 32 & ``I don't think they affected my opinions of the real world''  \\ \hline
       Real World Problem(s) helped broaden perspectives of course concepts. & 8 & 2 & ``These problems revealed how foundational concepts like logic and reasoning underpin various fields and real-life situations  broadening my perspective on their significance.'' \\ \hline
       Real World Problem(s) helped them acquired new skills. & 19 & 0 &  ``I can see more ways to solve and prove real-world problems I couldn't have solved before.''\\ \hline
       Real World Problem(s) helped them understand the course concepts better. & 108 & 134 & ``They helped me grasp concepts because I was able to see them in application and not just abstraction which helped a lot.'' \\ \hline
       Real World Problem(s) made course content feel more connected to real world. & 5 & 11 & ``It allowed us to realize how the concepts we learn in lecture can be used in real life.'' \\ \hline
       Real World Problem(s) made it clear why they were learning the content. & 18 & 19 &  ``Made me understand why learning all these abstract concepts matters.'' \\ \hline
       Real World Problem(s) made it more clear how course concepts could be applied to real life. & 3 & 5 & ``It was cool to see how we could take what we learned and put them into use.'' \\ \hline
       Real World Problem(s) were interesting or engaging. & 136 & 73 & ``I thought that they were a unique way to present the class concepts and made the homeworks more engaging as a result.'' \\ \hline
       Realize that problems in the real world are usually harder to deal with. & 2 & 0 & ``The real world is a little more complicated than I thought.'' \\ \hline
       Reflection about the ambiguity of the English language. & 7 & 0 & ``English is inherently an inexact language and a lot of the grammar we use in real life is confusing.'' \\ \hline
       Saw course concepts in future courses and/or careers. & 40 & 9 & ``This question was the first real-world homework question that directly applied to computer/program related problems  which in my opinion  was where I truly felt that learning and understanding the concepts of logic or discrete math is directly or indirectly helpful for my CS future.'' \\ \hline
       Saw course concepts in the real world. & 336 & 319 & ``They made the concepts in this class more like a part of a real world and not just concepts on paper.'' \\ \hline
       Spent significant amount of time on the Real World Problem(s). & 11 & 50 & ``It took me more time than the rest of the homework combined!'' \\ \hline
       Struggled to find a real world example for Real World Problem(s). & 21 & 46 & ``I struggled to find my own example so I had to pick from the given choices.'' \\ \hline
       Unclear about what DM is at the beginning of the quarter.& 4 & 0 & ``unclear on what discrete math actually is'' \\ \hline
       Unclear about what TCS is at the beginning of the quarter.& 1 & 0 & ``I don't quite know what theoretical computer science and discrete math is quite yet'' \\ \hline
       Unsure about real life relevance of course concepts at the beginning of the course. & 1 & 0 & ``in real life  I can't quite tell whether or not it will be relevant.'' \\ \hline
       Unsure what DM applications there are in their future career. & 0 & 1 & ``311 was a very interesting course. I just don't see it widely used in real life. Logic exists yes  but it is not used strictly or enforced a lot in the real world. The most I see it used in CS is \textbar \textbar and \&\&. I've spoken to many SWE's who have been in the field from 3 years all the way to 30 years  and when they saw my logic homework they tell me they have never used it before in their careers haha. It starts to get me wondering if logic is truly as important to computer science as colleges make it seem.'' \\ \hline
       Viewed course content as important or useful. & 32 & 78 & ``I kind of just see the importance of why we are learning what we are learning because they are indeed pretty useful in real life.'' \\ \hline
       Wanted future classes to struggle thought Real World Problem(s). & 2 & 1 & ``I found them hard its only fair the next class has to go through it too.'' \\ \hline
       Wanted Real World Problem(s) to have a single provided example. & 0 & 1 & ``I would've liked a more concrete provided example instead of letting us pick our own examples just so we could've had a more in-depth exploration of the topic.'' \\ \hline
       Were excited for the class at the beginning of the quarter. & 18 & 13 & ``I am excited to take this class.'' \\ \hline
       Worried about getting open-ended questions of Real World Problem(s) right. & 11 & 15 & ``I was not really as confident that what I was doing was correct'' \\ \hline
    \caption{\centering All codes from intervention quarter's qualitative feedback.}
    \label{tab:codebookI}
\end{longtable}

\begin{table}
    \centering
    \begin{tabular}{|p{2in}|c|c|p{2in}|}
    \hline
       \textbf{Code}  & $n$ \textbf{in DM} & $n$ \textbf{in PS} & \textbf{Example Quote}\\ \hline 
       Believe that DM is foundational. & 2 & 0 & ``discrete math is foundational knowledge'' \\ \hline
       Excited about the course at the beginning of the course. & 7 & 3 & ``I am excited to take this class.'' \\ \hline
       Felt positive about course. & 15 & 1 & ``Liked this class!'' \\ \hline
       Saw course concepts in the real world. & 1 & 0 & ``Knowing how the structure of the sentences in English changes the meaning entirely is helpful.'' \\ \hline
       Unsure about relevance of course concepts to real world or future career. & 5 & 0 & ``I'm unsure at this time of what the practical applications to development are of discrete math.'' \\ \hline
       Wished the course had more applications of course concepts. & 1 & 0 & ``Wish we dived more into the theoretical concepts of mathematics in relation to computer science. Glossed over it in last week when it felt like the entire basis for the course and future courses.'' \\ \hline
    \end{tabular}
    \caption{All codes from control quarter's qualitative feedback.}
    \label{tab:codebookC}
\end{table}

\end{document}